\documentstyle[preprint,epsfig,aps]{revtex}

\long\def\Omit#1{}

\tightenlines
\begin{document}

\draft
%% \preprint{}

\title{Magnetic moment and radius of the nucleon in a nonlocal model of
meson-nucleon interaction  }

\author{A. Yu. Korchin\footnote{e-mail address: alex$\_$korchin@rbcmail.ru}}
\address{National Science Center 'Kharkov Institute of Physics and
Technology', 61108 Kharkov, Ukraine}

%% \date{\today}
\maketitle

\begin{abstract}

Magnetic moment and radius of the nucleon are calculated in a nonlocal
extension of the chiral linear $\sigma -$model.  Properties of the
nonlocal model under the vector and axial transformations are
considered. The conserved electromagnetic and vector currents,
and partially conserved axial vector current are obtained. In the
calculation of the nucleon electromagnetic vertex the $\pi- $ and
$\sigma -$loop diagrams are included. Contribution from vector mesons
is added in the vector meson dominance model with a gauge-invariant
photon-meson coupling.  The nonlocality parameter associated with the
$\pi N$ interaction is fixed from the experimental magnetic moment of
the neutron.  Other parameters (nonlocality parameter for the $\sigma
N$ interaction and the mass of the $\sigma -$meson) are constrained by
the magnetic moment of the proton.  The calculated electric and
magnetic mean-square radii of the proton and neutron are in
satisfactory agreement with experiment.
\end{abstract}

%%% \bigskip \noindent

%% \\ {\em Keywords}:  magnetic moments of proton and neutron;
%% electric and magnetic mean-square radii;
%% nonlocal meson-nucleon interaction; linear sigma-model}

%%%%%%%%%%%%%%%%%%%%%%%%%%%%%%%%%%%%%%%%%%%%%%%%%

\section{Introduction}
\label{sec:introduction}

Calculation of the nucleon magnetic moment has a long story. In fifties
Fried~\cite{Fri} calculated pion-loop contribution to the
electromagnetic (EM) form factors of the neutron. The authors
of~\cite{Ern} pointed out to the importance of the Ward-Takahashi (WT)
identity as a requirement for any model for the $\gamma NN$ vertex.
Recoil corrections to the fixed-source static model~\cite{Wal59} were
calculated in~\cite{Ern} in a model with an extended pion-nucleon
interaction and pseudo-vector (PV) coupling. However agreement between
the theory and experimental values of the magnetic moments of the
proton ($\mu _{p}$) and neutron ($\mu _{n}$) was poor.

The authors of~\cite{Nau,Tie} evaluated the anomalous magnetic moments
(AMM)s of the proton, $\kappa _{p}=\mu _{p}-1$, and the neutron,
$\kappa _{n}=\mu _{n}$, while studying the $\gamma NN$ vertex of the
off-mass-shell (bound) nucleon. Naus and Koch~\cite{Nau} calculated the
off-shell EM form factors in the pion-loop approximation using the
pseudo-scalar (PS) $\pi N$ interaction.  The off-shell effects were
shown to be appreciable, although the calculated AMMs, $\kappa
_{p}=0.51,\,$ $\kappa _{n}=-3.7$, did not agree with the experimental
values $\kappa _{p}^{\exp }\approx 1.793$ and $\kappa _{n}^{\exp
}\approx -1.913$\footnote{Anomalous magnetic moments are given
hereafter in units of nuclear magneton $ e\hbar /2m_N c$}.  A more
refined approach based on the vector meson dominance (VMD) was applied
in~\cite{Tie}. The pion-loop contribution and the $\pi NN$ form
factors were included. To account for the gauge invariance of the
$\gamma NN$ vertex the authors used a recipe from ref.~\cite{Gro87}.
It was shown~\cite{Tie} that the VMD vertex supplemented
with the pion-loop contribution leads to a better description of AMMs,
{\it vis.} $\kappa _{p}=1.9,$ $\kappa _{n}=-2.26$ for the PS, and
$\kappa _{p}=1.74,$ $\kappa _{n}=-2.08$ for the PV $\pi N$ coupling. In
a dynamical model~\cite{Sch95} the effects of the $\Delta -$resonance
and VMD\ were investigated in both the space-like and time-like regions
of the photon momentum.  An extended version of the VMD model was
applied, which lead to results different from the conventional VMD
model used in~\cite{Tie}. The calculated AMMs were $ \kappa _{p}=1.45$
and $\kappa _{n}=-1.65$. Recently a non-perturbative approach for the
$\gamma NN$ vertex has been developed in~\cite{Kon00}, where the
infinite number of the pion loops was included. The magnetic form factor at
the photon point was normalized to experiment.

In ref.~\cite{Dav93} the chiral linear $\sigma -$model~\cite{Gel60} has been
applied in calculation of the AMM of the proton.
Among other results it has been shown that the proton
AMM can be reproduced with a heavy $\sigma -$meson ($m_{\sigma }\approx 800$
MeV). As follows from our calculation (sect.~\ref{sec:results} below) the
neutron AMM turns out to be far off the experimental value. Thus it does not
seem possible to describe simultaneously the proton and neutron AMMs in the
linear $\sigma -$model, at least on the one-loop level.

In the present paper we develop a nonlocal extension of the linear $\sigma
-$model, and apply it in calculation of the EM properties of the proton and
neutron. We assume that the $\pi N$ ($\sigma N)$ interaction is governed
by the form function $H_{\pi }(x^{\prime }-x,x^{\prime \prime }-x)$ ($%
H_{\sigma }(x^{\prime }-x,x^{\prime \prime }-x))$, where the space-time
coordinates $x^{\prime },x^{\prime \prime }\ $ stand for the nucleon, and $x$
for the meson. The nucleon mass is as usual generated through a non-zero
vacuum expectation value of the $\sigma -$field, this leads to a
constrained form function $ H_{\sigma }(x^{\prime }-x,x^{\prime \prime }-x)
= \delta ^{4}(x^{\prime} -x^{\prime \prime }) h_{\sigma}(x^{\prime }-x) $. \
The similar form is used for the $\pi N$ function.
The EM interaction in the model is included by making
use of the minimal substitution in the so-called shift operators. This
procedure generates, in addition to the nucleon and pion EM vertices, a
sea-gull-like $NN\pi \gamma $ term.
Properties of the nonlocal model under the $SU(2)$ axial and vector
transformations are also different from properties of the local $\sigma -$%
model. We construct the conserved vector current and partially
conserved axial current by applying the corresponding minimal substitutions
in the action. These currents get contributions from the meson-nucleon
interaction similarly to the EM current.

 Further we calculate the $\gamma NN$ vertex in the covariant
perturbation theory in
the pion- and sigma-loop approximation. The cut-off in momentum space in the
Fourier transforms of the functions $h_{\pi ,\sigma }(x^{\prime }-x)$
ensures convergence of the loop integrals. Consistency of the calculation is
verified by checking the WT identity. We concentrate on the magnetic moments
and mean-square radii (MSR) of the proton and neutron and study
contributions from various
diagrams. For reasonable values of the nonlocality parameters and the
$\sigma -$meson mass we find agreement between the calculation and
experiment for AMMs. At the same time the $\pi-$ and $\sigma-$loop
contributions turn out to be insufficient to describe the observed MSRs. For
this reason the contributions coming from the vector mesons are added in
the version~\cite{Sch95} of the VMD model. In this version the photon
couples to the vector mesons via a gauge-invariant Lagrangian. Inclusion of
the $\rho -$ and $\omega -$mesons improves considerably description of the
electric and magnetic MSRs of the proton and neutron.

The paper is organized as follows. In sect.~\ref{sec:model} the nonlocal
model is introduced. The conserved EM current is constructed. Properties of
the model with respect to the axial and vector transformations are discussed
and the corresponding currents are obtained. In sect.~\ref{sec:magnetic} the
nucleon EM form factors are calculated in the one-loop approximation.
Expressions for AMMs are derived.  Lagrangian of the vector mesons is specified
and their contributions to the EM form factors and MSRs are obtained.
One important aspect of renormalization of the EM form factors is
discussed. Results of the calculation of
AMMs and MSRs, and discussion are presented in sect.~\ref{sec:results}.
Conclusions are given in sect.~\ref{sec:Conclusions}.
Appendix~\ref{app:equations} contains equations of motion in the nonlocal model.
The Ward-Takahashi identity for the EM vertex is verified in
Appendix~\ref{app:WT-identity}. Finally, Appendix~\ref{app:loop-integral}
includes details of calculation of the loop integrals and expressions for the
EM form factors, MSRs and the nucleon self-energy operator.

\vspace*{2mm}

\section{ Description of the model}
\label{sec:model}

Let us consider the following action for the nucleon ($N$),
pion ($\vec{\pi}$), and sigma ($\phi $) fields
\begin{eqnarray}
S =\int \{ \bar{N}(x) i\partial \hspace{-0.5em}/N(x)\,+ \frac{1}{2}%
\,(\partial^\mu \vec{\pi}(x) )^2 + \frac{1}{2} (\partial^\mu \phi (x))^{2}
-V(\phi ,\vec{\pi}) \} \mbox{\rm d}x + S_{int} \, ,
 \label{action1}
\end{eqnarray}
\begin{eqnarray}
S_{int} = -g\int \bar{N}(x^{\prime })[\phi (x)H_{\sigma }{(x^{\prime
}-x,x^{\prime \prime }-x)}\,
&+& i{\gamma_5  \vec{\tau} \vec{\pi}(x)} H_{\pi }%
{(x^{\prime }-x,x^{\prime \prime }-x)]}     \nonumber \\
&& \quad \quad \quad \quad  \quad \quad \quad \quad
 \times N(x^{\prime \prime })
\ \mbox{\rm d}x \mbox{\rm d}x^{\prime} \mbox{\rm d} x^{\prime \prime }  \,,
\label{eq:interaction1}
\end{eqnarray}
where $g$ is the coupling constant, $ \mbox{\rm d}x =\mbox{\rm d}^{4}x,$
$\partial ^{\mu }=\partial
/\partial x_{\mu }$ ,\  ${\partial \hspace{-0.5em}/=\gamma }_{\mu }\partial
^{\mu }$ and
\begin{equation}
V(\phi ,\vec{\pi})=\frac{\lambda }{4}[\phi (x)^{2}+{{{\vec{\pi}}(x)}}%
^{2}-\xi ^{2}]^{2}-c\phi (x) \,
\label{potential1}
\end{equation}
is the meson potential.
Eq.~(\ref{eq:interaction1}) is a nonlocal extension
of the meson-nucleon interaction in the chiral $SU(2)_{L}\times SU(2)_{R}$
linear $\sigma -$model with explicit symmetry-breaking term \ $c\phi $
(see, {\it e.g.}, \cite{Itz}, Ch.11, sect.11.4.1,
or~\cite{DeA}, Ch.5, sect.2.6).
The three-point form functions
$H _{\pi,\sigma}(x^{\prime }-x,x^{\prime \prime }-x)$
due to the translational invariance depend on the two four-vectors.
After imposing the
Lorentz invariance they become functions of $s_{1}=(x^{\prime }-x)^{2}$, $%
s_{2}=(x^{\prime \prime }-x)^{2}$ and $s_{12}=(x^{\prime }-x^{\prime \prime
})^{2}$. In general, $H_{\pi, \sigma}(x^{\prime }-x,x^{\prime \prime }-x) $
should fall off at $|s_{1}|$, $|s_{2}|,$ $|s_{12}|\
\rightarrow \ \infty $ and have correct local limit. At this point we
refer to~\cite{Pau53,Chr53} where different aspects of the
nonlocal $\bar{\psi} \psi \phi $ model~\cite{Kri52}
for the nucleon and neutral meson fields were considered. A review of some
nonlocal theories can be found in monographs~\cite{Efi77}.

In the same way as in the local model we
take $\phi (x)=\langle \phi \rangle +\sigma (x)$ and require the minimum
of $V(\langle \phi \rangle ,\langle \vec{\pi} \rangle )$ to be at $\langle
\phi \rangle =f_{\pi }$ and $\langle \vec{\pi}\rangle =0,$ where $f_{\pi }$
is the $\pi^{+} \rightarrow \mu^{+} \nu _{\mu }$ \ weak-decay
constant\footnote{Its experimental value is
$f_\pi^{exp}=92.4(2)$ MeV~\cite{Tim99}}.
This gives \ $\xi ^{2}=f_{\pi }^{2}- {c}/ \lambda f_{\pi}  $ \ and
the action takes the form
\begin{eqnarray}
S=\int \{\bar{N}(x)(i\partial \hspace{-0.5em}/-m_{N})N(x)\,
+ \frac{1}{2}[
(\partial ^{\mu }\vec{\pi}(x))^{2}-m_{\pi }^{2}] &+&
\frac{1}{2}[(\partial^{\mu }\sigma (x))^{2}-m_{\sigma }^{2}]
\nonumber \\
&&-V^{\prime }(\sigma {,\vec{\pi})}\}\ \mbox{\rm d}x
+S_{int}^{\prime }\,,
\label{action2}
\end{eqnarray}
\begin{eqnarray}
V^{\prime }(\sigma ,\vec{\pi}) = \lambda  [\sigma (x)^{2}
+\vec{\pi} (x)^{2} ]   \{ f_\pi \sigma(x) +\frac{1}{4}  [\sigma (x)^{2}
+\vec{\pi} (x)^{2} ]  \}   + \mbox{\rm const} \,,
\label{eq:potential2}
\end{eqnarray}
and $S_{int}^{\prime}$ is obtained from $S_{int}$ in
eq.~(\ref{eq:interaction1}) after replacing $\phi(x)$ by $\sigma(x)$.
The masses of the pion, sigma and nucleon are defined respectively by
\begin{equation}
m_{\pi }^{2}=\frac{c}{f_{\pi }},\;\ \ \ \ \ \ \ \ \ \ \ \ \ \ \ m_{\sigma
}^{2}=2\lambda f_{\pi }^{2}+\frac{c}{f_{\pi }}\,,\;\ \ \ \ \ \ \;\;\ \ \ \ \
\ \ \ m_{N}=gf_{\pi }\ F_{\sigma }
  \label{masses}
\end{equation}
with a constant $F_{\sigma }.$ In order to reproduce the
nucleon mass term the following condition
\begin{equation}
gf_{\pi }\int {\bar{N}(x^{\prime })N(x^{\prime
\prime }) H_{\sigma }{(x^{\prime }-x,x^{\prime \prime }-x)}\,
\mbox{\rm d}x \mbox{\rm d}x^{\prime} \mbox{\rm d}x^{\prime \prime }}
=  m_{N}\int \bar{N}(x) N(x) \mbox{\rm d}x
\label{eq:constr1}
\end{equation}
has been imposed. From eq.~(\ref{eq:constr1}) one obtains the constraint
\begin{equation}
\int H{_{\sigma }\ {(x^{\prime }-x,x^{\prime \prime }-x)}\, \mbox{\rm d}x
=\delta^4 (x}^{\prime }-x^{\prime \prime })F_{\sigma }\,.
\label{eq:constr2}
\end{equation}
We further introduce the Fourier transforms
\begin{equation}
 \hat{H}_{\pi ,\sigma }(p^{\prime },p^{\prime \prime })= \int {\exp }[{{%
-i(p^{\prime }\cdot y}}^{\prime }{-p}^{\prime \prime }{\cdot y}^{\prime
\prime }{{)]}\ } {H}{_{\pi ,\sigma }(y^{\prime },y}^{\prime \prime }{%
)\ \mbox{\rm d} y}^{\prime }{\mbox{\rm d} y}^{\prime \prime } \, ,
\label{Fourier1}
\end{equation}
with notation $a\cdot b\equiv a_{\nu }b^{\nu }$.
Then eqs.~(\ref{eq:constr2}) and (\ref{Fourier1})
 lead to the constraint in the momentum space $\hat{H}_{\sigma
}(p,p)=F_{\sigma }$ for any $p$. Hence $\hat{H}_{\sigma }(p^{\prime
},p^{\prime \prime })$ should depend on $p^{\prime }-p^{\prime \prime}$,
or $\hat{H}_{\sigma
}(p^{\prime },p^{\prime \prime })= \hat{h}_{\sigma }(p^{\prime }-p^{\prime
\prime }).$ We will also choose \ $\hat{H}_{\pi}(p^{\prime },p^{\prime
\prime }) = \hat{h}_{\pi }(p^{\prime }-p^{\prime \prime }) $ for the
pion-nucleon interaction.
In configuration space one has
$H_{\pi, \sigma}(x^{\prime
}-x,x^{\prime \prime }-x)= \delta^{4} (x^{\prime }-x^{\prime \prime })h_{\pi, \sigma
}(x^{\prime }-x)$, where
\begin{equation}
{h}_{\pi ,\sigma } (y) = (2 \pi)^{-4} \int {\exp } (i k \cdot y)
\hat{h}_{\pi ,\sigma }(k)\ \mbox{\rm d}^{4} k  \,
\label{Norm-ff}
\end{equation}
with $\hat{h}_{\sigma }(0) = F_{\sigma }$ due to eq.~(\ref{eq:constr2}).
The normalization of the form factors $\hat{h}_{\pi, \sigma} (k)$ at $k = 0$
is specified in subsect.~\ref{sec:one-loop}.
The constrained form function $h_{\pi} (x^{\prime}- x) $ or
$h_{\sigma} (x^{\prime}- x) $ describes an
extended interaction between the meson and the nucleon source, whereas due to
$\delta^{4} (x^{\prime }-x^{\prime \prime })$ there is no
internal coupling of the nucleon to itself.

The interaction (\ref{eq:interaction1}) can be rewritten for convenience as follows
\begin{eqnarray}
S_{int} &=&-g\int [\phi {{(x)}\,} \tilde{\rho }_{S}{(x)+} \vec{\pi} {(x)}%
\tilde{\vec{\rho}}{_{A}{(x)] \mbox{\rm d}x} \,,}
 \label{interaction3} \\
\tilde{\rho }_{S} (x) &=& \int \rho _{S}{(x}^{\prime }{)h}_{\sigma
}(x^{\prime }-x) \mbox{\rm d}x^{\prime } \,, \,\ \ \ \ \ \ \
\rho _{S}(x^\prime ) = {\bar{N}(x^\prime )N(x^\prime )}\,,
 \label{eq:RHOS} \\
\tilde{\vec{\rho}}_{A} (x)
&=& \int \vec{\rho}_{A}{(x}^{\prime }{)h}_{\pi }(x^{\prime
}-x) \mbox{\rm d}x^{\prime } \,, \,\;\ \ \ \ \
\vec{\rho}_{A} (x^\prime )=\bar{N}(x^\prime ) i\gamma_{5} \vec{\tau} N(x^\prime )\,.
\label{eq:RHOA}
\end{eqnarray}

%%%%%%%%%%%%%%%%%%%%%%%%%%%%%%%%%%%%%%%%%%%%%%%%%%%%%%%%%

\subsection{Electromagnetic current}
\label{sec:EM-current}

In this subsection we discuss properties of the model under the gauge $%
U(1)$ transformation. Note first that variation of the action (\ref{action1})
with $S_{int}$ in eq.~(\ref{interaction3})
under arbitrary infinitesimal variation of the fields leads to
\begin{equation}
\delta S=\int \frac{\partial }{\partial x^{\mu }}[\bar{N}(x)i\gamma
^{\mu }\delta N(x)+\partial ^{\mu }\phi (x)\delta \phi (x)+\partial ^{\mu }%
\vec{\pi}(x)\delta \vec{\pi}(x)] \mbox{\rm d}x \,,
\label{eq:action-variation}
\end{equation}
where equations of motion have been used (see Appendix~\ref{app:equations}).
Let us take the transformation
\begin{eqnarray}
N &\rightarrow &N-i \epsilon \ e\hat{Q}_{p}N \, ,\;\ \ \ \ \ \ \ \   \bar{N}%
\rightarrow \bar{N} +i \epsilon \ \bar{N} e\hat{Q}_{p} \, ,
\nonumber \\
\vec{\pi} &\rightarrow &\vec{\pi}-i\epsilon \ e\hat{Q}_{\pi }\vec{\pi} \, ,\;\ \ \
\ \ \ \ \  \ \    \phi \rightarrow \phi  \label{eq:gauge-trans}
\end{eqnarray}
with the parameter $\epsilon.\ $ In eqs.~(\ref{eq:gauge-trans})
$e$ is the proton charge,
$\hat{Q}_{p}={\frac{1}{2}}(1+\tau _{3})$ and $\hat{Q}_{\pi }=t_{3}$,
where ${\vec{\tau}}$ \ and $\vec{t}$ are the isospin matrices for the nucleon and
the pion respectively. For the constant $\epsilon $ the action is clearly
invariant, {\it i.e.} $\delta S=0$,
and therefore the following integral analogue of the current conservation holds
\begin{equation}
\int \frac{\partial }{\partial x^{\mu }}[{J}_{N,em}^{\mu }(x)+J_{\pi
,em}^{\mu }(x)] \mbox{\rm d}x = 0 \, ,
\label{eq:EM-int}
\end{equation}
where the nucleon and the pion EM currents are defined as
$J_{N,em}^{\mu}(x)=e\bar{N}(x)\gamma ^{\mu }\hat{Q}_{p}N(x)$ and
$J_{\pi ,em}^{\mu}(x)=-ie \partial^{\mu} \vec{\pi} \hat{Q}_{\pi} \vec{\pi}
=e(\vec{\pi}\times \partial ^{\mu }\vec{\pi})_{3}$.
 Eq.~(\ref{eq:EM-int}) is in line with ref.~\cite{Rze53} where
 integral conservation laws were studied for a general nonlocal action, and with
ref.~\cite{Pau53} where the related issues were addressed, in particular
construction of the baryon charge and four-vector of the energy-momentum in
framework of the nonlocal model~\cite{Kri52}.

Despite eq.~(\ref{eq:EM-int}) the current  ${J}_{N,em}^{\mu }(x)+J_{\pi
,em}^{\mu }(x)$ is not locally conserved, {\it i.e.}, \ $\partial _{\mu }[{J}%
_{N,em}^{\mu }(x)+J_{\pi ,em}^{\mu }(x)]\neq 0.$
A conserved current can be constructed by making use of the minimal substitution
\begin{equation}
\partial _{x}^{\mu }\rightarrow \partial _{x}^{\mu }+ie\hat{Q}
A^{\mu }(x)\,,  \label{eq:minimal}
\end{equation}
where $A^{\mu }(x)\,$\ is the EM field, $\partial _{x}^{\mu } =
\partial /\partial x_{\mu }$ and $\hat{Q}=\hat{Q}_p$ or $\hat{Q}_\pi.$
The free nucleon and meson terms in the action (\ref{action1})
give rise to the currents \ $J_{N,em}^{\mu }(x)$\ and \ $J_{\pi ,em}^{\mu
}(x).$ In order to apply eq.~(\ref{eq:minimal}) to the interaction
(\ref{interaction3}) we represent the $\pi N$ piece of \ $S_{int}$ in the form
\begin{eqnarray}
S_{\pi N} &=& -g \int \int \vec{\pi} (x) \vec{\rho}_{A} (x+y)
 h_{\pi} (y) \ \mbox{\rm d}x \mbox{\rm d}y
\nonumber \\
&=&-g\int \int \vec{\pi}(x)[ \exp (y\cdot \partial_{x}) N(x)]^\dagger
\gamma_{0} i \gamma_{5} \vec{\tau} [\exp (y\cdot \partial_{x}) N(x)] h_{\pi} (y)
 \ \mbox{\rm d}x \mbox{\rm d}y  \, ,
\label{eq:min-form}
\end{eqnarray}
where the `shift' operator $\ {\exp (y\cdot \partial _{x})}$ acts on the
nucleon field as follows: $\exp (y\cdot \partial _{x})N(x)= N(x+y)$. After
substituting eq.~(\ref{eq:minimal}) in eq.~(\ref{eq:min-form}) one can use
the identity
\begin{eqnarray}
\exp (O_{1}+O_{2}) &=&\bar{{\cal P}}\exp [\,\int_{0}^{1}O_{2}(t)\, \mbox{\rm d}t\,]\,
\exp(O_{1}) \,, \\
O_{2}(t) &=&\exp (O_{1}t)O_{2}\exp (-O_{1}t)\,,
\nonumber
\end{eqnarray}
for any operators $O_{1}$ and $O_{2}$, where $\bar{{\cal P}}\exp [...]$ denotes
the exponential antiordered in the parameter $t$. Choosing $O_{1}=y\cdot {%
\partial _{x}}$ and $\ O_{2}=ie\hat{Q}_{p}y\cdot A(x)$ one obtains
\begin{eqnarray}
S_{\pi N}\rightarrow S_{\pi N}\{A\} &=&-g\int \int \vec{\pi}(x)
\bar{N}(x^\prime ) \, {\cal P} \exp [-ie\chi (x^{\prime },x)]
\, i\gamma _{5}\vec{\tau}
 \nonumber \\
&& \times \bar{{\cal P}} \exp [ie\chi (x^{\prime },x)]N (x^{\prime }) \
{h_{\pi } {(x^\prime -x)}} \ \mbox{\rm d}x \mbox{\rm d}x^{\prime } \,,
\label{eq:EM-full} \\
\chi (x^{\prime },x) &=&(x^{\prime }-x) \cdot \int_{0}^{1}\hat{Q}_{p}{%
A(x(1-t)+x}^{\prime }{t)\, \mbox{\rm d}t} \,
\nonumber
\end{eqnarray}
with ${\cal P} \exp [...]$ being the ordered in $t$ exponential.
The modified interaction $S_{\pi N}\{A\}$ is invariant under the
transformations (\ref{eq:gauge-trans}) with the $x-$dependent parameter $%
\epsilon (x) $, if the EM field transforms as $A^{\mu
}(x)\rightarrow A^{\mu }(x)+\partial ^{\mu }\epsilon (x) $. The $\sigma N$
term does not give rise to the EM interaction.

We should note that eq.~(\ref{eq:EM-full}) gives the EM interaction in all
orders in charge. In the first order in $e$ we obtain the current by taking
the functional derivative
\begin{eqnarray}
&&J_{\pi N,em}^{\mu }(x) =-\frac{\delta }{\delta A_{\mu }(x)}S_{\pi
N}\{A\}|_{A=0} \nonumber \\
&&= -eg \int \int [ \vec{\rho}_{A}(y) \times \vec{{\pi}} (z)]_3 \ (y-z)^{\mu
}\int_{0}^{1}\delta ^{4}{(x-z(1-t)-y}t)\, \mbox{\rm d}t
\ h_{\pi }(y-z)\ \mbox{\rm d}y \mbox{\rm d}z \,.
\label{eq:EM-current}
\end{eqnarray}
 Now direct calculation
shows that the total EM current is conserved, {\it i.e.},
\begin{equation}
\frac{\partial }{\partial x^{\mu }}[{J}_{N,em}^{\mu }(x)+J_{\pi ,em}^{\mu
}(x)+J_{\pi N,em}^{\mu }(x)]=0 \,.
\label{eq:local-conserv}
\end{equation}
In case of the local $\pi N$ interaction, $\ h_{\pi }(y-z) \rightarrow
\delta ^{4}(y -z)$, the current $J_{\pi N,em}^{\mu }(x)$ vanishes.

One should keep in mind that there is an ambiguity in construction of the
current $J_{\pi N,em}^{\mu }(x).$\ Additional gauge invariant terms appear
if the integral in eq.~(\ref{eq:EM-full}) along the straight line
\begin{equation}
\int_{0}^{1}(x^{\prime }-x)\cdot {A(x(1-t)+}x^{\prime }t{)\, \mbox{\rm d}t}%
=\int_{x}^{x^{\prime }}{A}^{\mu }\,{\mbox{\rm d}l}_{\mu }
\end{equation}
is replaced by the integral over an arbitrary contour connecting the
points $x^{\nu }$ and $x^{\prime \nu }.$
This was noticed long ago by Bloch~\cite{Blo52} when developing a
nonlocal theory. Our following
consideration is restricted to the straight-line integration.

Finally note that the minimal substitution in the shift operator was applied
earlier in~\cite{Kor88,Kor91}. In particular, in~\cite{Kor91} the EM current
for the two nucleons in the Bethe-Salpeter formalism  was constructed; there
also some additional gauge-invariant contributions, proportional to the EM
tensor $F^{\mu \nu }(x),$ were discussed.

%%%%%%%%%%%%%%%%%%%%%%%%%%%%%%%%%%%%%%%%%%%

\subsection{Axial and vector currents}
\label{sec:axial+isospin}

Properties of the model under the isospin transformation with the parameters
$\epsilon ^{a}\;\;(a=1,2,3)$ \
\begin{eqnarray}
N &\rightarrow & N- i\frac{\vec{\tau}}{2} \vec{\epsilon}\ N \, , \;\ \ \
\ \ \ \ \ \ \          \bar{N}\rightarrow \bar{N} +\bar{N} \ i \frac{%
\vec{\tau}}{2} \vec{\epsilon} \,,  \nonumber \\
\ \vec{\pi}  &\rightarrow &\vec{\pi}  +\vec{\epsilon} \times \vec{\pi} \, ,
\; \ \ \ \ \ \ \ \ \ \ \ \ \  \phi \rightarrow \phi \,.
\label{eq:isospin-rotate}
\end{eqnarray}
are similar to the case considered in the previous subsection.  It is
easy to show from eq.~(\ref{eq:action-variation}) that the integral conservation law
\begin{equation}
\int \frac{\partial }{\partial x^{\mu }}[\vec{J}_{N,vec}^{\mu }(x)+\vec{J}%
_{\pi,vec}^{\mu }(x)] \mbox{\rm d}x = 0
\label{eq:vec-int}
\end{equation}
is fulfilled, where
$\vec{J}_{N,vec}^{\mu }(x)=\bar{N}(x)\gamma ^{\mu }\frac{%
\vec{\tau}}{2}N(x)$ and $\ \vec{J}_{\pi,vec}^{\mu }(x)=\vec{\pi}(x)\times
\partial ^{\mu }\vec{\pi}(x)$ are the conventional nucleon and pion
vector currents.

The axial $SU(2)$ transformation in the nonlocal $\sigma-$model is a more
interesting case because the action is not invariant under the
transformation. Consider the following variation of the fields
\begin{eqnarray}
N  &\rightarrow &N   -i \gamma _{5}\frac{\vec{\tau}}{2} \vec{\epsilon} \ N \, ,
\;\ \ \ \ \ \ \bar{N} \rightarrow \bar{N}  -\bar{N} \ %
i \gamma _{5} \frac{\vec{\tau}}{2}  \vec{\epsilon} \,,
\nonumber \\
\ \vec{\pi}  &\rightarrow &\vec{\pi} +\vec{\epsilon} \phi \, , \;\ \ \ \
\ \ \ \ \ \ \ \ \ \ \ \phi  \rightarrow \phi -\vec{\epsilon} \vec{\pi} \,.
\label{eq:chiral-rotate}
\end{eqnarray}
For the constant parameters $\epsilon^a$ it follows
from eq.~(\ref{eq:action-variation}) that
\begin{equation}
\delta S=\vec{\epsilon}\int \frac{\partial }{\partial x^{\mu }}[\vec{J}%
_{N,ax}^{\mu }(x)+\vec{J}_{M,ax}^{\mu }(x)] \mbox{\rm d}x \, ,
\label{eq:action-var-2}
\end{equation}
where the nucleon and the meson axial currents are $\vec{J}_{N,ax}^{\mu }(x)=
\bar{N}(x)\gamma ^{\mu }\gamma _{5}\frac{\vec{\tau}}{2}N(x)$ and $\vec{J%
}_{M,ax}^{\mu }(x)=\partial ^{\mu }\vec{\pi}(x)\phi (x)-\partial ^{\mu }\phi
(x)\vec{\pi}(x)$ respectively.
The same variation $\delta S$ can be calculated directly by
making use of $\ \delta \rho_S (x)= - \vec{\epsilon} \vec{\rho}_A (x)\ $
and $\ \delta \vec{\rho}_A (x)= \vec{\epsilon} \rho_S (x).\ $ Then
comparison of the result with eq.~(\ref{eq:action-var-2})
yields the equation
\begin{eqnarray}
&&\int \frac{\partial }{\partial x^{\mu }}[\vec{J}_{N,ax}^{\mu }(x)+\vec{J}%
_{M,ax}^{\mu }(x)] \mbox{\rm d}x = -c\int \vec{\pi}(x) \mbox{\rm d}x
\nonumber \\
&&\;\ \ \ \ \ \ \ \ \ \ \ \ \ \ \ \ \ \ \ \ +g\int \int [\rho _{S}{(x}%
^{\prime }{)\vec{\pi}(x)+\vec{\rho}}_{A}(x^{\prime })\phi (x)][{h}_{\sigma
}(x^{\prime }-x)-{h}_{\pi }(x^{\prime }-x)] \mbox{\rm d}x \mbox{\rm d}x^{\prime } \,.
 \label{eq:ax-int}
\end{eqnarray}
The first term on the r.h.s. of eq.~(\ref{eq:ax-int}) comes from the
symmetry-breaking term $c \phi $ leading to the finite pion mass,
while the second term is related to the difference between the form functions
for the $\sigma N$ and $\pi N$ interactions.

Note that eqs.~(\ref{eq:vec-int}) and (\ref{eq:ax-int}) are the integral
relations from which the corresponding local relations do not follow.
In order to find the axial and vector currents satisfying the local relations
let us use the following `minimal' substitutions
\begin{eqnarray}
\partial ^{\mu }N(x) &\rightarrow &\{\partial ^{\mu } +i \frac{\vec{\tau} }{2}%
[\gamma _{5}\vec{a}^{\mu }(x)+\vec{\rho}^{\mu }(x)] \}N(x) \, ,
\nonumber \\
\partial ^{\mu }\vec{\pi}(x) &\rightarrow &\partial ^{\mu }\vec{\pi}(x)-\phi
(x)\vec{a}^{\mu }(x)+\vec{\pi}(x)\times \vec{\rho}^{\mu }(x) \, ,
\nonumber \\
\partial ^{\mu }\phi (x) &\rightarrow &\partial ^{\mu }\phi (x)+\vec{\pi}(x)%
\vec{a}^{\mu }(x) \, ,
\label{eq:minimal-ax}
\end{eqnarray}
where the axial vector field $\vec{a}^{\mu }(x)$ and the vector
field $\vec{\rho}^{\mu}(x)$ are introduced.
They transform according to (\cite{DeA}, Ch.5, sect.4.3)
\begin{equation}
\vec{\rho} ^{\mu } \rightarrow \vec{\rho} ^{\mu }
+\vec{\epsilon} \times \vec{\rho} ^{\mu }  +\partial
^{\mu }\vec{\epsilon} \,, \;\;\;\;\;\;\;\;\;\;\;\;\;\;
\vec{a}^{\mu } \rightarrow
\vec{a}^{\mu } +\vec{\epsilon} \times \vec{a}^{\mu }
\end{equation}
under the isospin rotations, and
\begin{equation}
\vec{\rho} ^{\mu } \rightarrow \vec{\rho} ^{\mu }
+\vec{\epsilon} \times \vec{a}^{\mu } \,, \;\;\;\;\;\;\;\;\;\;\;\;\;\;
\vec{a}^{\mu } \rightarrow \vec{a}^{\mu }
+\vec{\epsilon}  \times
\vec{\rho} ^{\mu } +\partial ^{\mu } \vec{\epsilon}
\end{equation}
under the chiral rotations described by the $x-$dependent
parameters $\epsilon^a (x)$. The axial and vector currents can be obtained by
taking the functional derivatives
\begin{equation}
\vec{J}_{ax}^{\mu }(x)=-\frac{\delta }{\delta \vec{a}_{\mu }(x)}S\{a,\rho
\}|_{a=\rho =0} \, , \;\ \ \ \ \ \ \ \
\vec{J}_{vec}^{\mu }(x)=-\frac{\delta }{\delta \vec{\rho}_{\mu }(x)}
S\{a,\rho \}|_{a=\rho =0} \, .
\label{eq:f-der}
\end{equation}
Applying eqs.~(\ref{eq:minimal-ax}) to the free nucleon and meson terms in
the action gives the axial current $\vec{J}_{N,ax}^{\mu }(x)+\vec{J}%
_{M,ax}^{\mu }(x)$ and the vector current $\vec{J}_{N,vec}^{\mu }(x)+%
\vec{J}_{\pi,vec}^{\mu }(x)$ defined above
after eqs.~(\ref{eq:action-var-2}) and (\ref{eq:vec-int}).
Those coincide with the currents in the local $\sigma -$model
(\cite{Itz}, Ch.11, sect.11.4.1).
Additional contributions arise from the nonlocal interaction.
Proceeding similarly to the derivation of the EM interaction in
subsect.~\ref{sec:EM-current} we obtain
\begin{eqnarray}
S_{int} \rightarrow S_{int}\{ a, \rho \} =
&-& g\int \int {\bar{N}(x}^{\prime }{)}%
{{\cal P}}\exp [-i{\varphi }(x^{\prime },x)+i\gamma _{5}{\psi }(x^{\prime },x)]
\{ \phi (x){{h_{\sigma }{(x}}}^{\prime }-x{)}
\nonumber \\
&+&i\gamma_{5}\vec{\tau}\vec{\pi}(x){{h_{\pi }{(x}}}^{\prime }-x{) \} }\bar{{\cal P}}%
\exp [i{\varphi }(x^{\prime },x)+i\gamma _{5}{\psi }(x^{\prime },x)] %
N(x^{\prime })\ \mbox{\rm d}x \mbox{\rm d}x^{\prime } \, ,
\label{eq:Sint-arho} \\
{\varphi }(x^{\prime },x) &=&\frac{1}{2}(x^{\prime }-x)\cdot \int_{0}^{1}%
\vec{\tau}{\vec{\rho}{(x(1-t)+x}^{\prime }{t)}\, \mbox{\rm d}t}\,, \;\ \ \ \ \ \ \ \
\nonumber \\
{\psi }(x^{\prime },x) &=&\frac{1}{2}(x^{\prime }-x)\cdot \int_{0}^{1}\vec{%
\tau}\vec{a}{(x(1-t)+x}^{\prime }{t)\, \mbox{\rm d}t \,.}
\nonumber
\end{eqnarray}

It is straightforward now to obtain the currents from eqs.~(\ref{eq:f-der})
and (\ref{eq:Sint-arho}). The axial vector current is
\begin{eqnarray}
\vec{J}_{MN,ax}^{\mu }(x)&=&-g\int \int [\rho _{S}(y)\vec{\pi}(z)h_{\pi }(y-z)-%
\vec{\rho}_{A}(y)\phi (z) h_{\sigma }(y-z) ]  (y-z)^\mu
\nonumber \\
&& \times \int_{0}^{1}\delta ^{4}{(x-z(1-t)-y}t)\, \mbox{\rm d}t\ \mbox{\rm d}y
\mbox{\rm d}z \,,
\label{eq:MN-axial}
\end{eqnarray}
and the vector current is
\begin{equation}
\vec{J}_{\pi N,vec}^{\mu }(x)=-g\int \int [\vec{\rho}_{A}(y) \times \vec{\pi}(z)] \
(y-z)^{\mu }\int_{0}^{1}\delta ^{4}{(x-z(1-t)-y}t)\, \mbox{\rm d}t\ h_{\pi }(y-z)\
\mbox{\rm d}y \mbox{\rm d}z \,.
\label{eq:MN-vector}
\end{equation}
One can also show that
\begin{equation}
\int \frac{\partial }{\partial x^{\mu }}\vec{J}_{MN,ax}^{\mu }(x) \mbox{\rm d}x=\int
\frac{\partial }{\partial x^{\mu }}\vec{J}_{\pi N,vec}^{\mu }(x) \mbox{\rm d}x=0 \,,
\end{equation}
though the divergences of the currents do not vanish.
Eqs.~(\ref{eq:MN-axial}) and (\ref{eq:MN-vector})
are sufficient to obtain the locally conserved vector current and
partially conserved axial current that obey equations
\begin{eqnarray}
\frac{\partial }{\partial x^{\mu }}[\vec{J}_{N,vec}^{\mu }(x)+\vec{J}%
_{\pi,vec}^{\mu }(x)+\vec{J}_{\pi N,vec}^{\mu }(x)] &=&0 \,,
\label{eq:div-vector} \\
\frac{\partial }{\partial x^{\mu }}[\vec{J}_{N,ax}^{\mu }(x)+\vec{J}%
_{M,ax}^{\mu }(x)+\vec{J}_{MN,ax}^{\mu }(x)] &=&-c\vec{\pi}(x)+g\int [\rho
_{S}(x^{\prime })\vec{\pi}(x)
\nonumber \\
&&+\vec{\rho}_{A}(x^{\prime })\phi (x)][h_{\sigma }(x^{\prime }-x)-h_{\pi
}(x^{\prime }-x)] \mbox{\rm d}x^{\prime } \,,
\label{eq:div-axial}
\end{eqnarray}
where we can substitute $\phi (x)=f_{\pi }+\sigma (x)\ $\ and $c=f_{\pi
}m_{\pi }^{2}.$ If the form functions for the pion and sigma are
equal to each other then the axial current satisfies the simpler equation
\begin{equation}
\frac{\partial }{\partial x^{\mu }}[\vec{J}_{N,ax}^{\mu }(x)+\vec{J}%
_{M,ax}^{\mu }(x)+\vec{J}_{MN,ax}^{\mu }(x)]=-c\vec{\pi}(x)\,.
\end{equation}
Apparently the currents (\ref{eq:MN-axial}) and (\ref{eq:MN-vector})
vanish for the local $\pi N$ and $\sigma N$ interactions,
{\it i.e.}, if $h_{\pi} (y-z) \rightarrow \delta^4 (y-z)$,
$\ h_{\sigma} (y-z) \rightarrow \delta^4 (y-z)$. In this case
all equations of the local $\sigma -$model are restored.

In general case, taking the matrix element of eq.~(\ref{eq:div-axial}) at $%
x^\mu = 0$ between the one-pion state and the vacuum leads, in the lowest
order, to the PCAC relation
\begin{eqnarray}
&&\langle 0|\partial_x \cdot [{J}_{N,ax}^{ a }(0) +{J}_{M,ax}^{ a }(0) + {J}%
_{MN,ax}^{ a }(0) ]|\pi ^{b}(\vec{q})\rangle
\nonumber \\
&&= - c \delta ^{ab} +g [\hat{h}_{\sigma }{(0)-}\hat{h}_{\pi }{(0)}]\
\langle 0|\rho_S (0) |0\rangle \delta ^{ab} =-f_{\pi
}m_{\pi }^{2}\delta ^{ab}\,,
\label{eq:PCAC-pion}
\end{eqnarray}
where the pion states are normalized as follows:
$\langle \pi ^{a}(\vec{q}^{\prime })|\pi ^{b}(\vec{q})\rangle
=(2\pi )^{3} 2 \omega_q \delta^3 (\vec{q}-\vec{q}^{\prime }) \delta ^{ab}$ \ with
$\omega_q^{2} =\vec{q}^{2}+m_{\pi}^{2} $, and $a,b=1,2,3$. \
The term in eq.~(\ref{eq:PCAC-pion}) proportional to $g$
does not contribute because it involves the nucleon operators.
The zero vacuum expectation value of $\rho_S (0)$ can be
ensured by choosing the normal ordering of the nucleon operators in $S_{int}$.
Besides, one can normalize the form factors so that $\hat{h}_{\pi
}(0)=\hat{h}_{\sigma }(0)$.
It is seen that on the level of the matrix element
the PCAC relation has the same form as in the local $\sigma -$model~\cite{Gel60}.
On the operator level however there is an
additional term in eq.~(\ref{eq:div-axial}) proportional to the difference
between the nonlocality functions for the $\pi -$ and $\sigma -$ mesons.

The axial properties of the nucleon will be addressed in detail elsewhere.
Now we proceed to the calculation of the nucleon EM vertex.

%%%%%%%%%%%%%%%%%%%%%%%%%%%%%%%%%%%%%%%%%%%%%%%%%%%%%%%%%%%

\section{Magnetic moment and mean-square radius of the nucleon}
\label{sec:magnetic}

%%%%%%%%%%%%%%%%%%%%%%%%%%%%%%%%%%%%%%%%%%%%%%%%%%%%%%%%%%%%%

\subsection{Pion- and sigma-loop contributions to the electromagnetic vertex}
\label{sec:one-loop}

To calculate the nucleon EM vertex we start with the 3-point Green
function
\begin{eqnarray}
&&G_{\nu }(x_{2},x_{1};z) =\langle 0|TN(x_{2})\bar{N}(x_{1})
A_{\nu }(z)|0 \rangle
\nonumber \\
&=&(2\pi )^{-8}\int {\exp [i(p_{2}x_{2}-p_{1}x_{1}-q z )]\delta
^{4}(p_{1}+q-p_{2})G_{\nu }(p_{2},p_{1};q)\ \mbox{\rm d}^{4}p_{1}
\mbox{\rm d}^{4}p_{2} \mbox{\rm d}^{4}q} \, ,
\label{GreenF1}
\end{eqnarray}
where $T$ is the time-ordering operator.
The Fourier transform $G_{\nu }(p_{2},p_{1};q)$ of the Green function is
related to the irreducible $\gamma NN$ vertex function $\Gamma ^{\mu
}(p_{2},p_{1};q)$ via
\begin{equation}
G_{\nu }(p_{2},p_{1};q)=iS^{\prime }(p_{2})[-ie\Gamma ^{\mu
}(p_{2},p_{1};q)]iS^{\prime }(p_{1})[-iD_{\mu \nu }^{\prime }(q)]\,,
\label{GreenF2}
\end{equation}
where $S^{\prime }(p)$ ($D_{\mu \nu }^{\prime }(q)$) is the nucleon (photon)
dressed propagator, and $p_{2}=p_{1}+q$.
Next step is evaluation of the lowest order contributions to $\Gamma ^{\mu
}(p_{2},p_{1};q)$ in the perturbation theory. We take into account the pion- and
sigma-loop diagrams (see fig.~\ref{Fig.2} `a,b,c' and `d$\pm$')
which include contributions of the order $g^{2}.$
The proton and neutron vertex functions read
\begin{eqnarray}
&&\Gamma^{\mu }_p (p_{2},p_{1};q) =\gamma ^{\mu }-\Gamma _{a}^{\mu }+{\frac{1}{2}}
\Gamma _{b}^{\mu }+\Gamma _{c}^{\mu }+\Gamma _{d+}^{\mu }\, ,
\label{Gamma-p} \\
&&\Gamma^{\mu }_n (p_{2},p_{1};q) =\Gamma _{a}^{\mu }+\Gamma _{b}^{\mu }+\Gamma
_{d-}^{\mu }\,,
\label{Gamma-n}
\end{eqnarray}
where $\Gamma _{a}^{\mu }$ is the $\pi -$loop term with the photon coupled
to the pion, $\Gamma _{b}^{\mu }$ ($\Gamma _{c}^{\mu }$) is the $\pi -$loop (%
$\sigma -$loop) term with the photon attached to the proton:
\begin{eqnarray}
\Gamma _{a}^{\mu }&=& C\int {\Delta }_{\pi }{(k)\Delta _{\pi }}{%
(k-q)(2k-q)^{\mu }\hat{h}}_{\pi }{(k)\gamma }_{5}{S(p_{2}-k){\gamma }_{5}%
\hat{h}_{\pi }}{(q-k)\ \mbox{\rm d}^{4}k}\,,
\label{Gamma-a}  \\
\Gamma _{b}^{\mu }&=& -C\int {\Delta }_{\pi }{(k)\hat{h}_{\pi }}{(k){\gamma }%
_{5}S(p_{2}-k)\gamma ^{\mu }S(p_{1}-k){\gamma }_{5}\hat{h}_{\pi }}{(-k)\
\mbox{\rm d}^{4}k%
}\,,
\label{Gamma-b} \\
\Gamma _{c}^{\mu }&=& \frac{1}{2}C\int {\Delta }_{\sigma }{(k)\hat{h}_{\sigma }}%
{(k)S(p_{2}-k)\gamma ^{\mu }S(p_{1}-k)\hat{h}_{\sigma }}{(-k)\ \mbox{\rm d}^{4}k}\,,
\label{Gamma-c}
\end{eqnarray}
$\Delta _{\pi ,\sigma }(k)=(k^{2}-m_{\pi ,\sigma }^{2}+i0)^{-1}$ ($S(p)=({p%
\hspace{-0.5em}/}-m_N +i0)^{-1}$) is the free meson (nucleon) propagator with
physical value of the meson (nucleon) mass, and $C\equiv 2ig^{2}(2\pi )^{-4}$.
The contributions coming from $J_{\pi N,em}^{\mu}$
in eq.~(\ref{eq:EM-current}) are
\begin{eqnarray}
\Gamma _{d+}^{\mu } = -\Gamma _{d-}^{\mu }
&=&C \int \Delta_{\pi}(k) \,
\{  \int_{0}^{1}  \frac{\partial}{\partial k_\mu }        \,
\hat{h}_{\pi }(qt+k) \, \mbox{\rm d}t \, \gamma _{5} S(p_1 -k) \gamma _{5}
\,\hat{h}_{\pi } (-k)
\nonumber \\
&& +\hat{h}_{\pi }(k) \, \gamma _{5} S(p_2 -k) \gamma _{5}
\int_{0}^{1} \frac{\partial}{\partial k_\mu } \,
\hat{h}_{\pi }(qt-k) \, \mbox{\rm d}t \, \} \ \mbox{\rm d}^{4}k \,.
\label{c+-1}
\end{eqnarray}
The term $\Gamma_{d+}^{\mu }$ ($\Gamma _{d-}^{\mu }$) corresponds
to the $\pi ^{+}$ ($\pi^-$) in the intermediate state for the proton (neutron)
vertex. Gauge invariance of the $\gamma pp$ and $\gamma nn$ vertices is verified in
Appendix~\ref{app:WT-identity}.

The analytical form of the nonlocal form factors is chosen as follows
\begin{equation}
\hat{h}_{\pi ,\sigma }(k)={\frac{\lambda _{\pi ,\sigma }^{2}}{{\lambda _{\pi
,\sigma }^{2}-k}^{2}}}\,,  \label{analytic}
\end{equation}
where $\lambda _{\pi}$ and $\lambda_{\sigma }$ are the cut-off momenta.
The functions \ $%
\hat{h}_{\pi ,\sigma }(k)$ are normalized to unity at $k^{2}=0,$ \ thus \
$F_\sigma = \hat{h}_{\sigma} (0) = 1.$\ The
local $\pi N$ ($\sigma N$) interaction can be obtained by taking $\lambda _{\pi
}(\lambda _{\sigma })\rightarrow \infty$. Note that the covariant
parametrization is used though it leads to a singularity for the time-like
\ $k^{2}.$ This singularity may not be consistent with unitarity, however
this shortcoming can be important only at high energies. Note that sometimes
in the OBE models~\cite{Mac87} non-covariant parametrizations like $\
\lambda _{\pi }^{2}/({\lambda _{\pi }^{2}+}\vec{k}^{2}{)}\ $ are used for
the \ $\pi NN$ form factor.

Eqs.~(\ref{Gamma-a}) - (\ref{c+-1}) allow one to
calculate the $\gamma NN$ \ vertex for a general case of the off-mass-shell
(bound) nucleon. Here we restrict ourselves to the free nucleon with $\
p_{1}^{2}=p_{2}^{2}=m_N^{2}$. The corresponding vertex
is (\cite{Itz}, Ch.7, sect.7.1.3)
\begin{equation}
\bar{u}(p_{2})\Gamma ^{\mu }(p_{2},p_{1};q)u(p_{1})=\bar{u}%
(p_{2})[F_{1}(q^{2})\gamma ^{\mu }+i{\frac{\sigma ^{\mu \nu }q_{\nu }}{{2m_N}%
}}F_{2}(q^{2})]u(p_{1})\,,  \label{on-shell}
\end{equation}
where $\sigma ^{\mu \nu }=\frac{i}{2} [\gamma ^{\mu },\gamma ^{\nu }]$, \
$F_1 (q^2)$ and $F_2 (q^2)$ are the Dirac
and Pauli form factors normalized at the photon point as follows
\begin{equation}
F_{1}^p(0)=1\,, \quad \;    F_{1}^n (0)=0 \,,
 \quad \quad \quad \quad  \quad
F_{2}^p (0)=\kappa _{p}\,, \quad \;   F_{2}^n (0)=\kappa _{n}\,.
\label{F12norm}
\end{equation}
It is convenient to use the Sachs electric and magnetic form factors~\cite{Ern}
\begin{equation}
G_{E}(q^{2})=F_{1}(q^{2})+\frac{q^{2}}{4m_{N}^{2}}F_{2}(q^{2}) \, ,\;\ \ \ \ \ \
\ \ \ \ \ \ \ \ \ \ G_{M}(q^{2})=F_{1}(q^{2})+F_{2}(q^{2}) \, ,
\label{eq:Sachs}
\end{equation}
in terms of which the electric and magnetic MSRs are defined as~\cite{Eri88}
\begin{equation}
\langle r_{E}^{2} \rangle =6 G_{E}^{\prime} (0) =
6 [ F_{1}^\prime (0) +\frac{ F_2 (0) }{4m_N^2} ]  \,,\;\ \ \ \ \ \ \ \ \ \ \
\ \ \  \langle r_{M}^{2} \rangle =\frac{6}{\mu _{N}}
G_{M}^{\prime}(0)
\label{eq:radius-def}
\end{equation}
with  $\mu _{N}=G_M (0) = 1+\kappa _{p}$ ($\kappa _{n}$) for the proton
(neutron) and $G^{\prime}(0) \equiv d G(q^{2}) / dq^2 $ at $q^{2}=0$.

Calculation of the pion- and sigma-loop contributions to the EM form factors is
straightforward but cumbersome and we refer to
Appendix~\ref{app:loop-integral} where details are
collected. Here we only present results for AMMs
\begin{eqnarray}
\kappa _{p}&=&-\kappa _{a}+\frac{1}{2}\kappa _{b}+\kappa _{c}+\kappa _{d+}\,,
\label{Eq:prot-neutr-kappa1} \\
\kappa _{n}&=&\kappa _{a}+\kappa _{b}+\kappa _{d-} \,,
\label{Eq:prot-neutr-kappa}
\end{eqnarray}
where
\begin{eqnarray}
&&\kappa _{a}=-\frac{g^{2}m_N^{2} \lambda_\pi^4}{4\pi ^{2} l_\pi^4}
\int_{0}^{1}
\{\frac{2}{l_{\pi }^{2}} \ln \frac{D(m_{\pi })}{D(\lambda _{\pi })}%
+(1-\alpha )\big[\frac{1}{D(m_{\pi })}+\frac{1}{D(\lambda _{\pi })}\big]%
\,\}\alpha ^{2} \mbox{\rm d}\alpha \,,
\nonumber \\
&&\kappa _{b}=-\frac{g^{2}m_N^{2}\lambda_{\pi}^4 }{4\pi ^{2}}
\int_{0}^{1}\frac{\alpha ^{3}(1-\alpha )^{2}}{D(m_{\pi })D(\lambda
_{\pi })^{2}} \mbox{\rm d}\alpha \,,
\nonumber \\
&&\kappa _{c}=\frac{g^{2}m_N^{2} \lambda_{\sigma}^4   }{8\pi ^{2}}
\int_{0}^{1}\frac{\alpha ^{2}(2-\alpha )(1-\alpha )^{2}}{%
D(m_{\sigma })D(\lambda _{\sigma })^{2}} \mbox{\rm d}\alpha \,,  \label{abc}
\end{eqnarray}
\begin{equation}
\kappa _{d+}=-\kappa _{d-}=-{\frac{g^{2}m_N^{2} \lambda_\pi^4}{4\pi ^{2} l_\pi^4}}
\int_{0}^{1}\{%
\frac{2}{ l_{\pi }^{2} } \ln \frac{D(m_{\pi })}{D(\lambda
_{\pi })}+(1-\alpha )\big[\frac{3}{D(\lambda _{\pi })}-\frac{D(m_{\pi })}{%
D(\lambda _{\pi })^{2}}\big]\,\}\alpha ^{2} \mbox{\rm d}\alpha \,.
\label{cont+-}
\end{equation}
Here $l_{\pi, \sigma}^2 \equiv \lambda_{\pi, \sigma}^2 - m_{\pi, \sigma}^2 $ and
\begin{eqnarray}
D(m_{\pi ,\sigma }) = m_N^{2}\alpha ^{2}+m_{\pi ,\sigma }^{2}(1-\alpha
)\,,\;\;\;\;\;\;\ \ \ \ \ \ \ \;
D(\lambda _{\pi ,\sigma }) = m_N^{2}\alpha
^{2}+\lambda _{\pi ,\sigma }^{2}(1-\alpha ) \,.
\label{eq:Dml}
\end{eqnarray}

In the limit $\lambda _{\pi ,\sigma }\rightarrow \infty $ \ one reproduces
results in the local $\sigma -$model
\begin{eqnarray}
\kappa _{a} &=&-\frac{g^{2}m_N^{2}}{4\pi ^{2}}\int_{0}^{1}\frac{\alpha
^{2}(1-\alpha )}{D(m_{\pi })} \mbox{\rm d}\alpha \,,
\;\;\;\;\;\ \;\ \ \ \ \ \ \kappa
_{b}=-\frac{g^{2}m_N^{2}}{4\pi ^{2}}\int_{0}^{1}\frac{\alpha ^{3}}{D(m_{\pi })}%
\mbox{\rm d}\alpha \,,  \nonumber \\
\kappa _{c} &=&\frac{g^{2}m_N^{2}}{8\pi ^{2}}\int_{0}^{1}\frac{\alpha
^{2}(2-\alpha )}{D(m_{\sigma })} \mbox{\rm d}\alpha \,,
\;\;\;\;\;\;\;\ \ \ \ \ \ \ \
\kappa _{d+}=\kappa _{d-}=0\ .  \label{local-limit}
\end{eqnarray}

The expressions for the loop contributions
$\langle r_{E}^{2} \rangle _{\pi ,\sigma }$
and $\langle r_{M}^{2} \rangle _{\pi ,\sigma }$
to the nucleon electric and magnetic MSRs are given
in Appendix~\ref{app:loop-integral}.

%%%%%%%%%%%%%%%%%%%%%%%%%%%%%%%%%%%%%%%%%%%%%%%%%%%%%%

\subsection{Vector meson contributions to the electromagnetic vertex}
\label{sec:vector mesons}

We add the following Lagrangian density~\cite{Sch95} corresponding to the
vector mesons
\begin{eqnarray}
{\cal L}_{\omega ,\rho }(x) = \sum_{V=\omega ,\rho } \{
-  g_{V NN} \bar{N} ( \gamma^\mu
V_\mu  + \frac{\kappa _{V}}{4m_{N}}\sigma ^{\mu \nu }V_{\mu \nu } )  I_V  N
 &-&\frac{e}{2g_{V\gamma }}F^{\mu \nu }  V_{\mu
\nu }  \nonumber \\   &&
+ \frac{1}{2} m_{V}^{2}V^\mu  V_\mu  -\frac{1}{4}V^{\mu \nu }
V_{\mu \nu }   \} \,,
\end{eqnarray}
where  $V^{\mu }$ stands for the $\omega ^{\mu } $ or  ${\rho} ^{\mu },$ \
$ g_{VNN}$ is the (vector) coupling constant of the meson-nucleon interaction,
$\kappa _{V}$ is the ratio of tensor and vector couplings, \
$g_{V\gamma }$ is the photon-meson coupling constant, and the isospin factor
$I_V =1$ ($\tau_3$) for the $\omega$ ($\rho$).
Only the neutral $\rho^0 -$meson is included as it couples to the photon.
Furthermore, $F^{\mu \nu} = {\partial}^{\mu} A^\nu  - {\partial}^{\nu} A^\mu $
is the EM tensor and $V _{\mu \nu } =\partial_{\mu }V _{\nu }
-\partial _{\nu }V _{\mu }.$ The $\rho - \omega$ mixing is neglected.
This Lagrangian gives rise to the
nucleon EM vertex \ (fig.~\ref{Fig.2}, diagrams `e')
\begin{eqnarray}
\Gamma^\mu (p_{2},p_{1};q)_{\omega ,\rho }
&=&\frac{g_{\omega NN}}{g_{\omega \gamma }}\frac{q^{2}}{m_{\omega
}^{2}-q^{2} -im_{\omega } \Gamma _{\omega } (q^2 ) }(%
\tilde{\gamma}^{\mu}+i\frac{\sigma ^{\mu \nu }q_{\nu }}{2m_{N}}\kappa _{\omega })
\nonumber \\
&+& \tau _{3}\frac{g_{\rho NN}}{g_{\rho \gamma }}\frac{q^{2}}{m_{\rho
}^{2}-q^{2} -im_{\rho }\Gamma _{\rho } (q^2) }(\tilde{\gamma}^{\mu }+i\frac{\sigma
^{\mu \nu }q_{\nu }}{2m_{N}}\kappa _{\rho }) \, .
\label{eq:VMD-vertex}
\end{eqnarray}
To avoid problems with the gauge invariance for the
off-mass-shell nucleons one can introduce~\cite{Sch95} $
\tilde{\gamma}^{\mu }\equiv \gamma ^{\mu }-q \hspace{-0.5em}/ q^{\mu }/
q^{2}$ (there is no pole for real photons due to the $q^{2}$ factor in
eq.~(\ref{eq:VMD-vertex})). This modification is however irrelevant
for the on-shell nucleons since the
additional term does not contribute. The decay widths of the mesons,
$\Gamma _{\omega } (q^2 )$  and $\Gamma _{\rho} (q^2 )$,
can be omitted for the photon space-like momenta.

The contributions from the vector mesons to the EM form factors read
\begin{eqnarray}
F_{1}(q^{2})_{\omega ,\rho } &=&q^{2}(\frac{G_{\omega }}{m_{\omega
}^{2}-q^{2}}\pm \frac{G_{\rho }}{m_{\rho }^{2}-q^{2}}) \, ,  \nonumber \\
F_{2}(q^{2})_{\omega ,\rho } &=&q^{2}(\frac{G_{\omega }\kappa _{\omega }}{%
m_{\omega }^{2}-q^{2}}\pm \frac{G_{\rho }\kappa _{\rho }}{m_{\rho }^{2}-q^{2}%
})\,,
\label{eq:VMD-ff}
\end{eqnarray}
where `plus' stands for the proton, `minus' for the neutron, and
$G_{V}\equiv g_{VNN}/g_{V\gamma }$.
The form factors including all the diagrams in fig.~\ref{Fig.2} are
\begin{eqnarray}
&&F_{l}^p (q^{2}) = \delta_{l1} + F_{l}^p (q^{2})_{\pi, \sigma}
+F_{l}^p (q^{2})_{\omega, \rho} \,,
\nonumber  \\
&&F_{l}^n (q^{2}) = F_{l}^n (q^{2})_{\pi, \sigma} +F_{l}^n (q^{2})_{\omega, \rho}
 \,,
\label{eq:total-formfactors}
\end{eqnarray}
where $l =1,2$ and  $F_{l} (q^{2})_{\pi, \sigma}$ are given in eqs.~(\ref{Fln}).
There is no contribution from the vector
mesons to the nucleon magnetic moment, whereas they contribute to
the radii. From eqs.~(\ref{eq:VMD-ff}) and (\ref{eq:radius-def}) one
finds the corresponding electric and magnetic MSRs
\begin{equation}
\langle r_{E}^{2}\rangle_{\omega ,\rho }
=6(\frac{G_{\omega }}{m_{\omega }^{2}}\pm \frac{%
G_{\rho }}{m_{\rho }^{2}}) \, , \;\ \ \ \ \ \ \ \ \ \ \ \ \ \ \ \ \ \
\langle r_{M}^{2}\rangle_{\omega ,\rho }
=\frac{6}{\mu _{N}} \Big[ \frac{G_{\omega }(1+\kappa
_{\omega }) }{m_{\omega }^{2}}\pm
\frac{G_{\rho }(1+\kappa _{\rho })}{m_{\rho }^{2}} \Big] \,.
\label{eq:VMD-MSR}
\end{equation}

The meson-photon couplings are fixed from the $\omega \rightarrow e^{+}e^{-}$
and \ $\rho \rightarrow e^{+}e^{-}$ decay widths~\cite{Kli96}: \ $g_{\rho
\gamma }=5.03$ and \ $g_{\omega \gamma }=17.05$. These values approximately
follow the $SU(3)$ pattern $g_{\omega \gamma }/g_{\rho \gamma }=3.$ The
typical meson-nucleon couplings are collected in
table~\ref{tab:VMD}.
The values shown in the second row correspond to the so-called
universality of Sakurai (\cite{DeA}, Ch.5, sect.4) and the $SU(3)$ symmetry. The
universality requires \ $2g_{\rho NN}=g_{\rho \gamma }=\gamma _{\rho},$
where $\gamma_\rho$ is a universal coupling constant of the $\rho$ to any particle,
and the $SU(3)$ gives~\cite{Kli96}: $g_{\omega \gamma }=3g_{\rho \gamma }$
and $%
g_{\omega NN}=3g_{\rho NN}.$ We can use this approximation for an estimate.
Assuming equal masses $m_\rho = m_\omega = m_V = 770$ MeV we get the
radii squared induced by the vector mesons
\begin{eqnarray}
{\langle r_{E}^{2} \rangle}_{\omega ,\rho } &
=&\frac{6}{m_{V}^{2}}\approx 0.39\;\mbox{\rm fm}%
^{2}\,,
\nonumber \\
{\langle r_{M}^{2}\rangle}_{\omega ,\rho } &
=&\frac{6}{m_{V}^{2}(1+\kappa _{p})}(1+\frac{%
\kappa _{\omega }+\kappa _{\rho }}{2})\approx 0.40\;\mbox{\rm fm}^{2}
\end{eqnarray}
for the proton, and
\begin{eqnarray}
{\langle r_{E}^{2} \rangle}_{\omega ,\rho }&=&0 \, ,
\nonumber \\
{\langle r_{M}^{2}\rangle}_{\omega ,\rho }&
=&\frac{6}{m_{V}^{2}\kappa _{n}}(\frac{\kappa
_{\omega }-\kappa _{\rho }}{2})\approx 0.38\; \mbox{\rm fm}^{2}
\end{eqnarray}
for the neutron (values $\kappa _{\rho }=3.7,\, \kappa
_{\omega }=0.$ have been used). According to this estimate the $\omega$ and $%
\rho$ give rise to almost equal magnetic radii of the proton and neutron, as
well as the electric radius of the proton, though the values are well below
the experiment. In table~\ref{tab:VMD} the MSRs
${\langle r_{E/M}^{2} \rangle}_{\omega ,\rho }$
calculated with the other values of the $\rho N$ and $\omega N$ coupling
constants and physical masses of the mesons are also presented.
As it is seen, the MSRs depend on the coupling constants, although
the variations of the radius $R_{E}$ or $R_{M}$ do not
exceed 20\% (except $R_{En}^2$).

Adding the contribution from the $\sigma -$ and $\pi -$ meson loops we
obtain the total MSRs
\begin{equation}
\langle r_{E/M}^{2} \rangle =\langle r_{E/M}^{2} \rangle _{\pi ,\sigma }
+\langle r_{E/M}^{2} \rangle_{\omega ,\rho }
\label{eq:MSR-loop+VMD}
\end{equation}
for the proton and neutron. The calculated radii are compared with
 experiment in sect.~\ref{sec:results}.

%%%%%%%%%%%%%%%%%%%%%%%%%%%%%%%%%%%%%%%%%%%%%%%%%%%%%%%%%%%%%%

\subsection{Renormalization}        \label{sec:renorm}

In general, the $\gamma NN$ vertex needs a renormalization in accordance
with: $\Gamma _{R}^{\mu }(p_{2},p_{1};q)=Z_{1}\Gamma ^{\mu }(p_{2},p_{1};q),$
where $Z_{1}$ is vertex renormalization constant (\cite{Itz}, Ch.7, sect.7.1.3).
The renormalized vertex
(subscript `R' stands for renormalized quantities) obeys the
condition at $ p_2 \rightarrow p_1$: $\
\bar{u}(p_{1})\ \Gamma _{R}^{\mu }(p_{1},p_{1};0)\ u(p_{1})
=\bar{u}(p_{1})\ {%
\hat{Q}_{p}}\gamma ^{\mu }\ u(p_{1}).$ \
 Substituting the on-mass-shell vertex (\ref{on-shell})
in this equation yields $Z_{1}=1/F_{1}^{p}(0)$, where $F_1^{p} (0)$ is the
unrenormalized proton form factor at $q^2 = 0.$ Correspondingly, one has
$F_{1,2}(q^{2})_{R}=F_{1,2}(q^{2})/F_{1}^{p}(0)$. In literature
the two different renormalization schemes have been discussed: the
subtractive and the multiplicative ones. In the subtractive scheme~\cite
{Fri,Ern,Nau,Fle92} one expands $1/F_{1}^{p}(0)$ in powers of $g$ and
retains terms of the order $g^{2}$ in the one-loop calculation.
In the multiplicative scheme~\cite{Tie,Sch95} no
expansion is used. In table~\ref{tab:renorm} the renormalization rules
in the two schemes are compared.

As is seen from table~\ref{tab:renorm}, the renormalized AMM
$\kappa _{R}$ and the MSR  for the Dirac form factor
${\langle r_{1}^{2} \rangle}_R$  can be essentially
different in the two schemes if $Z_{1}$ differs considerably from unity.
Such a situation occurs in many calculations, {\it e.g.}, in \cite
{Ern,Tie,Sch95} where $Z_{1}$ is 0.3 -- 0.4 or less. In
the calculation below we follow the subtractive scheme as it consistently takes
into account terms of the same order. On the opposite, in the
multiplicative scheme some terms of the higher orders
may be artificially generated.

One test of the model is vanishing of the neutron form factor $%
F_{1}^{n}(q^{2}) $ at $q^{2}=0.$ This condition is fulfilled in our
calculation with the accuracy of $10^{-6}$. An important test of the gauge
invariance is fulfillment of the equation $Z_{1}=Z_{2},$ where $Z_{2}$ is
the wave-function renormalization constant. The latter is calculated
from the nucleon self-energy in Appendix~\ref{app:loop-integral}.
The equation $Z_{1}=Z_{2}$ is rather well satisfied
in the numerical calculation.

%%%%%%%%%%%%%%%%%%%%%%%%%%%%%%%%%%%%%%%%%%%%%%%%%%%%%%%%%%%%%%%

\section{Results of calculation and discussion}
\label{sec:results}

There are several parameters in the model: $\lambda _{\pi }$, $%
\lambda _{\sigma }$ and $m_{\sigma }$. There is also an ambiguity in value
of $g$ which in the local $\sigma -$model was pointed out in~\cite{Dav93}.
In view of a sensitivity of AMMs to $g$ this issue deserves attention.
In the $\sigma -$model the formula $m_{N}=gf_{\pi }$ is the Goldberger-Treiman
relation $r_{A}m_{N}=gf_{\pi }$ with the axial coupling
constant $r_{A}$ equal to unity (\cite{DeA}, Ch.5, sect.2.5).
The corresponding meson-nucleon coupling constant $g$
comes out rather small, namely $g=10.2$.
However, it is known that
renormalization of the axial coupling leads to the value $r_{A}^{exp} = $ 1.2573(28),
which increases $g$ to $12.8$. The latter number is close to the physical $%
\pi N$ coupling constant $\ g_{\pi NN}=13.01,$ which follows from the
modern value~\cite{Tim99}
 $\ f_{\pi NN}^{2}/4\pi =0.0745(6)$ through the relation $g_{\pi
NN}/2m_N =f_{\pi NN}/m_{\pi }$. The physical value is used in the
calculation below.

The pion cut-off momentum $\lambda _{\pi }$ can be fixed from
the neutron AMM because the latter does not depend on the sigma parameters.
In fig.~\ref{Fig.3} (left panel) the neutron AMM is shown as a
function of $\lambda _{\pi }.$ As it is seen, the experimental AMM is
reproduced with $%
\lambda _{\pi }=1.56$ GeV. In the local theory ($\lambda _{\pi }\rightarrow
\infty $) \ $\kappa _{n}$ turns out to be $-3.52$ which is too big. We note
that the sea-gull term $\kappa _{d-}$ is important, it contributes about
15\% to AMM at $\lambda _{\pi }=1.56$ GeV.

Having fixed the pion cut-off we calculate the proton AMM as a function of $%
m_{\sigma }$ (fig.~\ref{Fig.3}, right panel). In the figure $\lambda _{\sigma
}\rightarrow \infty $ is chosen. As it is seen the agreement with experiment is
observed for  $m_{\sigma }=915$ MeV. If we choose $\lambda _{\sigma }$ equal
to $\lambda _{\pi }$ then the $\sigma $ mass decreases to 290 MeV. The
results for AMMs and radii are presented in table~\ref{tab:results}. In
the calculation of MSRs we used the $\omega N$ and $\rho N$ couplings from
table~\ref{tab:VMD} (parameters from the third row corresponding to an effective
Lagrangian from~\cite{Noz90}).

The following comments regarding AMMs are in order. First, in the local
$\sigma -$model (second row in table~\ref{tab:results}) there is no agreement
with experiment for any value of $m_{\sigma }.$ It is possible to
get the proton AMM correct (with  $m_{\sigma }\approx 700$\ MeV), however it
is not possible to get both the proton and neutron AMMs correct at the same
time. This observation was one of the motivations for a nonlocal extension
of the $\sigma -$model.

Second, the variant with $\lambda _{\sigma }$ $=\lambda _{\pi }$ corresponds
to the chiral-symmetrical case. It requires however too low $\sigma -$meson
mass that does not look realistic.

Third, the set of parameters $\lambda _{\pi }=1.56$ GeV and\ $\lambda
_{\sigma }\rightarrow \infty $ \ (third row in table~\ref{tab:results})
implies that the $\sigma N$ interaction is local, as could be expected for
a heavy meson. The $\sigma -$meson can be viewed as an
approximation to various scalar-isoscalar exchanges ({\it e.g.},
two-pion, f$_{0}$(400--1200) or f$_{0}$(980)~\cite{PDG99}).
The light pion interacts with the nucleon via a nonlocal
Lagrangian;  the size of the nonlocality in configuration space is about
$\ {\langle r^2_{\pi N} \rangle}^{1/2}
= (\ 6 d\hat{h}_{\pi} (k) / dk^2 \ |_{k^2 =0} \ )^{1/2} =
\sqrt{6} \lambda _{\pi }^{-1} \approx 0.31$ fm.

As for the electric and magnetic radii, their sensitivity to
the $\pi $ and $\sigma$ cut-off parameters is not very strong (compare numbers
for $R_E^2$ or $R_M^2$ within a column in table~\ref{tab:results}).
For definiteness we discuss the calculation with
$\lambda _{\pi }=1.56$ GeV, $\lambda_{\sigma }\rightarrow \infty $
and $m_\sigma = 915$ MeV. In table~\ref{tab:radii}
the contributions from which the MSRs are formed
are shown separately.
It is seen that among the $\pi-$ and $\sigma-$loop terms the dominant
contribution to the electric radius comes from the diagram `a' in fig.~\ref{Fig.2},
where the photon couples to the virtual pion.
The $\sigma$ loop is also essential for the proton and contributes about 30\%,
the sea-gull diagrams `d+,\ d-' bring in
$\langle r_{E}^{2} \rangle _{\pi ,\sigma }$ about 10\%.
In the magnetic radius the diagrams  `a, b, c' (in fig.~\ref{Fig.2})
for the proton, and `a, b' for the neutron
are equally important.
It is worth mentioning that $\langle r_{E}^{2} \rangle _{\pi ,\sigma }$
is a sum of the contributions from the diagrams `a -- d',
while $\langle r_{M}^{2} \rangle _{\pi ,\sigma }$ is a more complicated
superposition of separate diagrams because both the numerator and denominator in
the definition (\ref{eq:radius-def}) of the magnetic radius are built up of
the diagrams `a -- d'.
The $\omega-$ and $\rho-$mesons contribute additively to the electric and
magnetic MSRs (see eq.~(\ref{eq:MSR-loop+VMD})) for
in the chosen VMD model  the
corresponding EM form factors (\ref{eq:VMD-ff}) vanish at $q^2 =0$.

It appears that the total loop contributions
$\langle r_{E/M}^{2} \rangle _{\pi ,\sigma }$ underestimate
noticeably the observed MSRs.
The vector mesons improve the agreement with experiment (compare the last
but one and the last columns in table~\ref{tab:radii}).
As it is seen, the electric $R_{Ep}$ (magnetic $R_{Mp}$)
radius of the proton is described with the
accuracy 3\% (7\%); the magnetic radius $R_{Mn}$ of the neutron differs from
experiment by about 5\%. The negative sign of
$R_{En}^{2}$ is reproduced, but not the magnitude.
Note that this calculation is performed with the one set of
the  $\rho N$ and $\omega N$ coupling constants.
Unfortunately there is a dependence of the radii on these parameters
({\it cf.} table~\ref{tab:VMD}).

%%%%%%%%%%%%%%%%%%%%%%%%%%%%%%%%%%%%%%%%%%%%%%%%%%%%%%%

\section{Conclusions}            \label{sec:Conclusions}

In the paper a nonlocal model for interacting nucleon, $\pi -$ and $\sigma
-$mesons has been developed. It is an extension of the chiral
linear sigma-model and allows for space-time form functions $h_{\pi,
\sigma}(x' - x)$ describing the $\pi N$ and $\sigma N$ interactions. While
these interactions are characterized by one coupling constant $g$
as in the local
model, the form functions can be different. The nonlocality may account for
degrees of freedom not included explicitly in the model.

The conserved electromagnetic
current has been obtained in the model via the minimal substitution in the
so-called shift operators. In the similar way the conserved vector
current and partially conserved axial current have been derived.
An important feature of the nonlocal model is that all these
currents get additional contributions from the meson-nucleon
interaction.

 The model has been applied in the calculation of the low-energy
electromagnetic properties of the nucleon, namely the magnetic moment and
the electric and magnetic mean-square radii. By varying the nonlocality
parameters of the $\pi N $ and $\sigma N$ interactions
($\lambda_\pi$ and $\lambda_\sigma$) and the sigma mass it
turns out possible to fit the magnetic moments of the proton and neutron. In
particular, the neutron magnetic moment allows one to fix $\lambda_\pi$,
while the proton magnetic moment gives at least two options for the
choice of $\lambda_\sigma$ and $m_\sigma$: a) local $\sigma N$ interaction
with $\lambda_\sigma \to \infty$ and heavy sigma-meson ($m_\sigma=915$ MeV),
b) chiral-symmetrical case with $\lambda_\sigma = \lambda_\pi$ and
light sigma-meson ($m_\sigma=290$ MeV).
In general, $\lambda_\pi =$1.56 GeV sets the upper limit of energies
where the model can be considered as an effective model.

At the same time the $\pi$ and $\sigma$ one-loop contributions are
not sufficient to describe the observed radii. The situation is improved if the
vector-meson contributions are added. For this purpose we have used the
version~\cite{Sch95} of the VMD model in which the photon coupling
to the vector mesons is described by a gauge-invariant Lagrangian.
The radii calculated based on the $\pi $, $\sigma $ and
$\omega $, $\rho $ contributions are in satisfactory
agreement with experiment for the proton and neutron.

In future we plan to study other observables in
low-energy nuclear/hadron physics in framework of this model.

%%%%%%%%%%%%%%%%%%%%%%%%%%%%%%%%%%%%%%%%%%%%%%%%%
\acknowledgments
%%%%%%%%%% \section*{Acknowledgments}
I would like to thank Prof. S.V. Peletminsky, Dr. V.D. Gershun and Dr. V.A. Soroka
for discussions.

%%%%%%%%%%%%%%%%%% APPENDIX %%%%%%%%%%%%%%%%%%%%%%%%%%%%%%%
%%%%%%%%%% \begin{appendix}
\appendix

\section{Equations of motion in the nonlocal model}
\label{app:equations}

The variational principle applied to eqs.~(\ref{action1}) and (\ref{interaction3}),
under the condition that variations of the fields vanish on the boundaries
of the integration region, leads to the equations of motion
\begin{eqnarray}
&&i\partial \hspace{-0.5em}/N(x)-g\int [\phi (x^{\prime })h_{\sigma
}(x-x^{\prime })+i\gamma _{5}\vec{\tau} \vec{\pi} (x^{\prime })
h_{\pi }(x-x^{\prime
})] \mbox{\rm d}x^{\prime }N(x) =0 \,,
\nonumber \\
&& \bar{N}(x) i \overleftarrow{\partial \hspace{-0.5em}/}
+g \bar{N}(x)\int [\phi
(x^{\prime })h_{\sigma }(x-x^{\prime })+i\gamma _{5}\vec{\tau} \vec{\pi} (x^{\prime
})h_{\pi }(x-x^{\prime })] \mbox{\rm d}x^{\prime } =0 \,,
\nonumber \\
&&\Box \vec{\pi} (x) +g \tilde{\vec{\rho}}_{A} (x) + \lambda \vec{\pi}
(x)[\phi (x)^{2}+ {{\vec{\pi}}(x)}^{2}-\xi ^{2}] =0 \,,
\nonumber \\
&& \Box \phi (x)+g \tilde{{\rho}}_{S}{{(x)}\,+}\lambda \phi
(x)[\phi (x)^{2}+ {{\vec{\pi}}(x)}^{2}-\xi ^{2}]-c = 0\,,
\nonumber     %%%%%%%\label{app:EOM}
\end{eqnarray}
with $\Box =\partial ^{\mu }\partial _{\mu }.$ The derivation of these
equations proceeds similarly to that in the $\bar{\psi} \psi \phi$ model of
Kristensen and M$o\hspace{-0.5em}/$ller~\cite{Kri52}.
After substitutions $\phi(x) = f_\pi +\sigma(x),$ $c=f_\pi m^2_{\pi}$ and
$\xi^2=f^2_{\pi}-m^2_{\pi}/\lambda$ (see sect.~\ref{sec:model})
in these equations
the mass terms for the nucleon, pion and sigma appear explicitly.
%%%%%%% \end{appendix}

%% \begin{appendix}
\section{Ward-Takahashi identity in one-loop approximation}
\label{app:WT-identity}

In this Appendix we check gauge invariance expressed in terms of the WT
identity. The identity reads (see, {\it e.g.}, \cite{Itz}, Ch.7, sect.7.1.3)
\begin{equation}
q_{\mu }\Gamma ^{\mu }(p_{2},p_{1};q)=\hat{Q}_{p}[S^{\prime
}(p_{2})^{-1}-S^{\prime }(p_{1})^{-1}]=\hat{Q}_{p}[{q\hspace{-0.5em}/}%
-\Sigma (p_{2})+\Sigma (p_{1})]\, ,
\label{eq:WT-identity}
\end{equation}
where $\Sigma (p)$ \ is the nucleon self-energy operator. In the order $%
g^{2}$ the latter can be written as
\begin{eqnarray}
\Sigma (p) =\Sigma _{\pi }(p)+\Sigma _{\sigma }(p)
&=& -{\frac{3}{2}}C \int \hat{{h}}%
_{\pi }(k)\gamma _{5}S(p-k)\gamma _{5}\hat{h}_{\pi }(-k)\Delta _{\pi
}(k)\, \mbox{\rm d}^{4}k
\nonumber \\&&
+\frac{1}{2}C \int \hat{h}_{\sigma }(k)S(p-k)\hat{h}_{\sigma }(-k)\Delta
_{\sigma }(k)\, \mbox{\rm d}^{4}k \, ,
\label{eq:self-energy}
\end{eqnarray}
where the factor $3$\ comes from the sum over charge states of the
intermediate pion.

First we check the WT identity for the neutron. One can use the identities
\begin{eqnarray}
S(p_{2}-k){q\hspace{-0.5em}/}S(p_{1}-k) &=& S(p_{1}-k)-S(p_{2}-k)\, ,
\label{Eq:ident0} \\
q\cdot (2k-q)\Delta _{\pi }(k)\Delta _{\pi }(k-q) &=& \Delta _{\pi
}(k-q)-\Delta _{\pi }(k)\, .
\label{Eq:ident1}
\end{eqnarray}
Contraction of $q_{\mu }$ with $\ \Gamma
_{d \pm }^{\mu }$\ is evaluated with the help of
\begin{eqnarray}
q_{\mu } \int_{0}^{1} \  \frac{\partial}{\partial k_\mu }
\hat{h}_{\pi } (qt+k) \, \mbox{\rm d}t
&=&\hat{h}_{\pi }(k+q) -\hat{h}_{\pi }(k)\, , \\
q_{\mu } \int_{0}^{1} \ \frac{\partial}{\partial k_\mu }
\hat{h}_{\pi } ( qt-k) \, \mbox{\rm d}t
&=& \hat{h}_{\pi }(-k) - \hat{h}_{\pi }(q-k)\, .
\label{Eq:ident2}
\end{eqnarray}
Collecting results for all terms in eq.~(\ref{Gamma-n}) one obtains
the required identity
\begin{eqnarray}
q_{\mu }(\Gamma _{a}^{\mu }+\Gamma _{b}^{\mu }+\Gamma _{d-}^{\mu })=0\, .
\nonumber
\end{eqnarray}

More care should be taken to verify the WT identity for the proton as the
r.h.s. of eq.~(\ref{eq:WT-identity}) is not zero.
Using eqs.~(\ref{Eq:ident0}) - (\ref{Eq:ident2}) we find
\begin{eqnarray}
q_{\mu }({\frac{1}{2}}\Gamma _{b}^{\mu }) &=&{\frac{1}{3}%
} [ \Sigma _{\pi }(p_{1})-\Sigma _{\pi }(p_{2})  ]\, ,  \nonumber \\
q_{\mu }(-\Gamma _{a}^{\mu }+\Gamma _{d+}^{\mu }) &=&{\frac{2}{3}} [ \Sigma
_{\pi }(p_{1})-\Sigma _{\pi }(p_{2})  ] \, ,  \nonumber \\
q_{\mu } \Gamma _{c}^{\mu } {} &=& {} \Sigma _{\sigma
}(p_{1})-\Sigma _{\sigma }(p_{2})\, .
\nonumber
\end{eqnarray}
These relations mean that the WT identity for the $\gamma pp$ vertex holds
separately for the contributions generated by the intermediate $\pi ^{0}$, $%
\pi ^{+}$, and the $\sigma $. The sum of the above equations is the WT identity
for the proton in  eq.~(\ref{eq:WT-identity}) (without the term
$q_\mu \gamma^\mu = q\hspace{-0.5em}/$).

%% \end{appendix}

%% \begin{appendix}

\section{Evaluation of loop integrals}
\label{app:loop-integral}

It is convenient first to use the relations that follow from eq.~(\ref
{analytic})
\begin{eqnarray}
\hat{h}_{\pi }(k)\Delta _{\pi }(k) &=&\frac{\lambda _{\pi }^{2}}{l_\pi^2}
\sum_{i=1,2}(-1)^{i+1}\frac{1}{k^{2}-m_{i}^{2}}\,,  \nonumber \\
\hat{h}_{\pi }^{2}(k)\Delta _{\pi }(k) &=& \frac{\lambda _{\pi }^{4}}{l_{\pi
}^{4}} \Big(1 - l_{\pi}^{2} \frac{\partial }{\partial \lambda _{\pi }^{2}} %
\Big)\sum_{i=1,2}(-1)^{i+1}\frac{1}{k^{2}-m_{i}^{2}}\,,  \nonumber
\end{eqnarray}
where $\ m_{1}=m_{\pi },\,m_{2}=\lambda _{\pi }$ and $l_\pi^2 \equiv
\lambda_\pi^2 - m_\pi^2;$ the similar formulas hold for the $\sigma -$meson.
Then all integrands can be written in terms of products of the three
multipliers in denominators. We apply the Feynman identity
\begin{eqnarray}
\frac{1}{ABC^{n+1}\,} &=&\frac{\Gamma (n+3)}{\Gamma (n+1)}%
\int_{0}^{1}\mbox{\rm d}\alpha _{1}\int_{0}^{1-\alpha _{1}}\mbox{\rm d}\alpha _{2}\frac{\alpha
_{3}^{n}}{(\alpha _{1}A+\alpha _{2}B+\alpha _{3}C)^{n+3}}\,,
\nonumber\\
\alpha _{3} &=&1-\alpha _{1}-\alpha _{2}\,.
\nonumber
\end{eqnarray}
Integration over $k$ is carried out using the
dimensional-regularization method (see, {\it e.g.,}~\cite{Col84}).
The following integrals for arbitrary four-momentum $Q^{\mu }$ and scalar $R$
are used
\begin{eqnarray}
\int \frac{[1,\,k^{\alpha },\,k^{\alpha }k^{\beta }]}{(k^{2}-2k\cdot
Q+R)^{n+3}} \mbox{\rm d}^{d}k
&=&\frac{i\pi ^{d/2}(-1)^{n+3}}{\Gamma (n+3)}\{\,\frac{%
\Gamma (n+3-d/2)}{(Q^{2}-R)^{n+3-d/2}}\ [1,\,Q^{\alpha },\,Q^{\alpha }Q^{\beta
}]  \nonumber \\
&&-\frac{\Gamma (n+2-d/2)}{2(Q^{2}-R)^{n+2-d/2}}\ [0,\,0,\,g^{\alpha \beta
}]\,\}\,,  \nonumber
\end{eqnarray}
%%%
where $\Gamma (z)$ is the Gamma-function and $g^{\alpha \beta}$ is the metric tensor.
While calculating the EM vertex for the on-shell nucleon we can drop terms
proportional to \ $q^{\mu }$ and $q{\hspace{-0.5em}/}$, and make use of the
Dirac equation for initial and final states $u(p_{1})$ and $\bar{u}(p_{2})$
in eq.~(\ref{on-shell}). The Gordon identity for the general
off-mass-shell case
\[
p_{1}^{\mu }+p_{2}^{\mu }=p\hspace{-0.5em}/_{2}\gamma ^{\mu }+\gamma ^{\mu }p%
\hspace{-0.5em}/_{1}-i\sigma ^{\mu \nu }q_{\nu }
\]
is useful while reducing formulas to eq.~(\ref{on-shell}).

\begin{figure}[1]
\begin{center}
%%% \leavevmode
\epsfig{figure=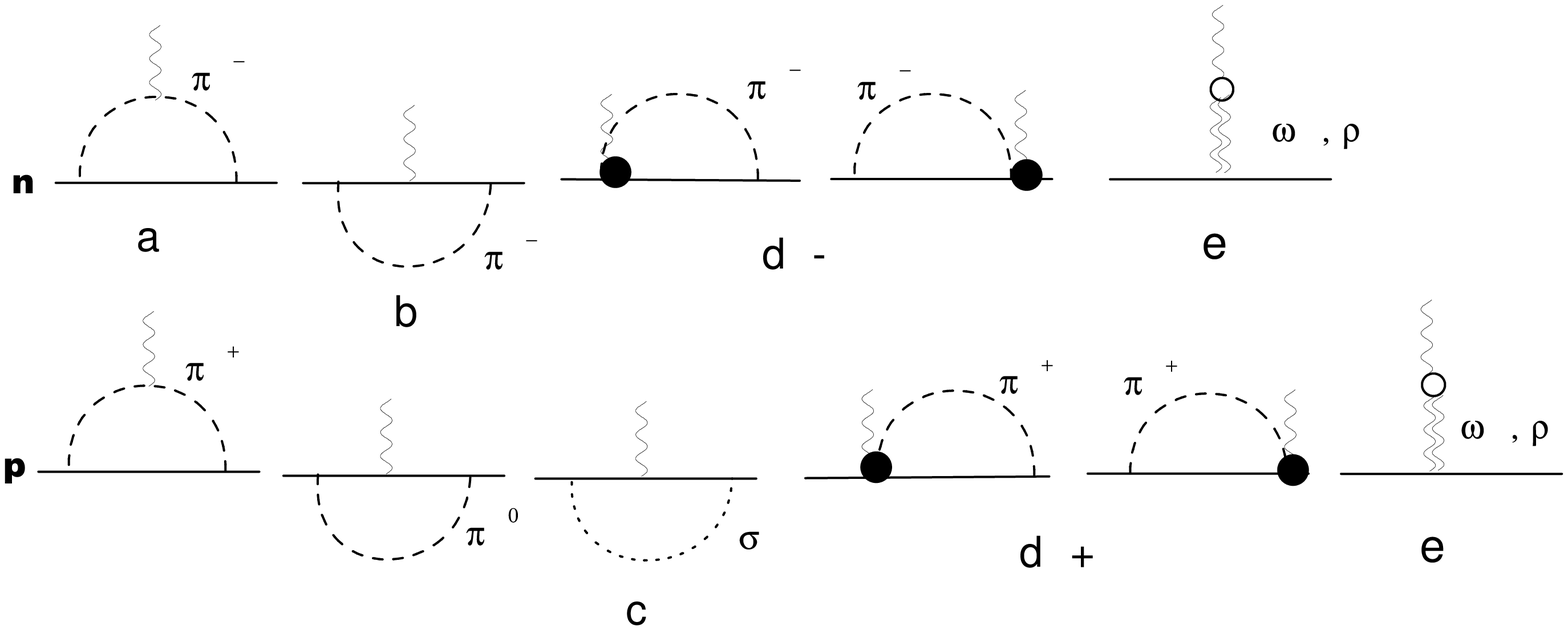,height=11.4cm}
\end{center}
\caption[Fig.2]{ Electromagnetic vertex of the nucleon. The first row of
diagrams shows the neutron vertex, the second row the proton vertex
(without the $\protect\gamma^{\mu}$ term).  Dashed
lines depict $\protect\pi -$meson, dotted lines $\protect\sigma -$meson,
wavy lines photon, double-wavy lines $\protect\omega -$ and  $\protect\rho -
$mesons, and solid lines  nucleon.}
\label{Fig.2}
\end{figure}

\begin{figure}[2]
\begin{center}
%%% \leavevmode
\epsfig{figure=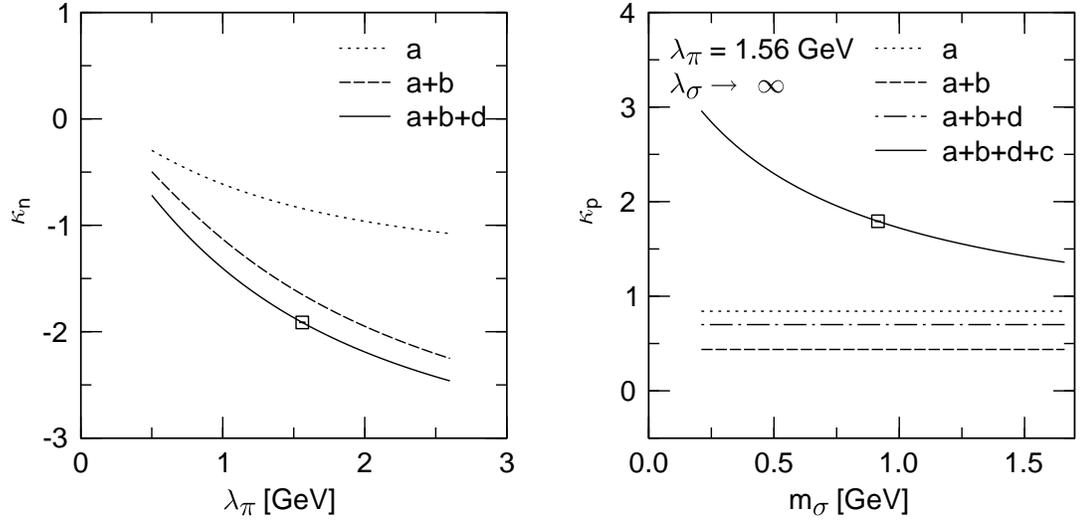,height=18cm}
\end{center}
\caption[Fig.3]{Left panel: the neutron anomalous magnetic moment as a
function of the pion cut-off momentum. Right panel: the proton anomalous
magnetic moment versus the mass of the $\protect\sigma-$meson. Experimental
values are indicated by the open squares. Various contributions in eqs.~(\ref
{Eq:prot-neutr-kappa1}) and (\ref{Eq:prot-neutr-kappa}) corresponding to
the diagrams `a,b,c' and `d$\pm$' in fig.~\ref{Fig.2} are also shown. }
\label{Fig.3}
\end{figure}

In this way we obtain the following contributions from the diagram `a' in
fig.~\ref{Fig.2}
\begin{eqnarray}
F_{1}(q^{2})_{a} &=&\frac{g^{2} \lambda_\pi^4}{4\pi ^{2}l_\pi^4 }
\sum_{i,j=1,2}(-1)^{i+j}%
\int_{0}^{1}\int_{0}^{1}\Big[ \frac{m_{N}^{2}\alpha^{2}}{ {\cal D}_{ij} } -\frac{\pi
^{d/2-2}\Gamma (2-d/2)}{2 {\cal D}_{ij}^{2-d/2}}\Big](1-\alpha)\
\mbox{\rm d}\alpha \mbox{\rm d}\beta  \,,
\label{F1a} \\
F_{2}(q^{2})_{a} &=&-\frac{g^{2}m_{N}^{2}\lambda_\pi^4 }{4\pi ^{2} l_\pi^4}
\sum_{i,j=1,2}(-1)^{i+j}\int_{0}^{1}\int_{0}^{1}
\frac{\alpha^2 (1-\alpha)}{ {\cal D}_{ij} } \
\mbox{\rm d}\alpha \mbox{\rm d}\beta \,,
\label{F2a}    \\
{\cal D}_{ij} &=&m_{N}^{2}\alpha^{2}+m_{i}^{2}(1-\alpha)
+(m_{j}^{2}-m_{i}^{2})(1-\alpha)\beta -q^{2}(1-\alpha)^{2}\beta (1-\beta)\,.
\nonumber
\end{eqnarray}
The contributions in fig.~\ref{Fig.2} `b' read
\begin{eqnarray}
F_{1}(q^{2})_{b} &=&\frac{g^{2} \lambda_\pi^4}{8\pi ^{2}\l_\pi^4}\Big(1 -l_{\pi }^{2}
\frac{\partial }{\partial \lambda _{\pi }^{2}} \Big)\sum_{i=1,2}(-1)^{i+1}%
\int_{0}^{1}\int_{0}^{1}
\Big[\frac{m_{N}^{2}\alpha^{2}+q^{2}\alpha^{2}\beta (1-\beta)}{{\cal D}_{i}}
\nonumber \\
&&- \frac{\pi ^{d/2-2} \Gamma (2-d/2)(2-d)}{2 {\cal D}_{i}^{2-d/2}}%
\Big] \alpha \ \mbox{\rm d}\alpha \mbox{\rm d}\beta \,,
\label{F1b} \\
F_{2}(q^{2})_{b} &=&-\frac{g^{2}m_{N}^{2} \lambda_\pi^4}{4\pi ^{2}l_\pi^4}
\Big(1-l_{\pi }^{2}
\frac{\partial }{\partial \lambda _{\pi }^{2}} \Big)\sum_{i=1,2}(-1)^{i+1}%
\int_{0}^{1}\int_{0}^{1}\frac{\alpha^{3} }{ {\cal D}_{i}} \
\mbox{\rm d}\alpha \mbox{\rm d}\beta \,,
\label{F2b}
\\
{\cal D}_{i} &=&m_{N}^{2}\alpha^{2}+m_{i}^{2}(1-\alpha)
-q^{2}\alpha^{2}\beta (1-\beta) \,.  \nonumber
\end{eqnarray}
Similarly, the $\sigma $ contribution to the proton form factors is
\begin{eqnarray}
F_{1}(q^{2})_{c} &=&\frac{g^{2} \lambda_\sigma^4}{16\pi ^{2}l_\sigma^4}
\Big(1 -l_{\sigma }^{2} \frac{%
\partial }{\partial \lambda _{\sigma }^{2}} \Big)\sum_{i=1,2}(-1)^{i+1}%
\int_{0}^{1}\int_{0}^{1}\Big[ \frac{m_{N}^{2}(2-\alpha)^{2}+q^{2}\alpha^{2}
\beta (1-\beta)}{%
 {\cal D}_{i}}  \nonumber \\
&&-\frac{ \pi ^{d/2-2} \Gamma (2-d/2)(2-d)}{2 {\cal D}_{i}^{2-d/2}}%
\Big] \alpha\ \mbox{\rm d}\alpha \mbox{\rm d}\beta \,,
\label{F1c} \\
F_{2}(q^{2})_{c} &=&\frac{g^{2}m_{N}^{2} \lambda_\sigma^4}{8\pi ^{2} l_\sigma^4}
\Big(1 -l_{\sigma }^{2}
\frac{\partial }{\partial \lambda _{\sigma }^{2}} \Big)\sum_{i=1,2}(-1)^{i+1}%
\int_{0}^{1}\int_{0}^{1}
\frac{\alpha^{2}(2-\alpha) }{{\cal D}_{i}}\ \mbox{\rm d}\alpha \mbox{\rm d}\beta \,,
\label{F2c} \\
{\cal D}_{i} &=&m_{N}^{2}\alpha^{2}+m_{i}^{2}(1-\alpha)
-q^{2}\alpha^{2} \beta (1-\beta)\,.  \nonumber
\end{eqnarray}
For the sea-gull diagram `d+' in fig.~\ref{Fig.2} we find
\begin{eqnarray}
F_{1}(q^{2})_{d+} &=&-\frac{g^{2} \lambda_{\pi }^{4}}{2\pi ^{2} l_\pi^2}%
\sum_{i=1,2}(-1)^{i+1}\int_{0}^{1} \mbox{\rm d}t \int_{0}^{1}\int_{0}^{1}\Big( \frac{%
m_{N}^{2}\alpha^{2}}{ {\cal D}_{i}^{2}}-\frac{1}{2 {\cal D}_{i}} \Big)
(1-\alpha)^{2}\beta\ \mbox{\rm d}\alpha \mbox{\rm d}\beta \,,
\label{F1d+} \\
F_{2}(q^{2})_{d+} &=&\frac{g^{2}m_{N}^{2} \lambda_{\pi }^{4}}{2\pi ^{2} l_\pi^2}%
\sum_{i=1,2}(-1)^{i+1}\int_{0}^{1} \mbox{\rm d}t \int_{0}^{1}\int_{0}^{1}\frac{%
\alpha^{2}(1-\alpha)^{2}\beta}{ {\cal D}_{i}^{2}}
\ \mbox{\rm d}\alpha \mbox{\rm d}\beta \,,
\label{F2d+} \\
{\cal D}_{i} &=&m_{N}^{2}\alpha^{2}+m_{i}^{2}(1-\alpha)+(\lambda _{\pi
}^{2}-m_{i}^{2})(1-\alpha)\beta + q^{2}t(1-\alpha)\beta
\{\alpha +t[(1-\alpha) \beta-1]\} \,,
\nonumber
\end{eqnarray}
and for the diagram `d-' in fig.~\ref{Fig.2}: $%
F_{1}(q^{2})_{d-}=-F_{1}(q^{2})_{d+},$ \ $%
F_{2}(q^{2})_{d-}=-F_{2}(q^{2})_{d+}.$ \ In these formulas $\ m_{i}=\{m_{\pi
},\,\lambda _{\pi }\}$ \ for \ the `a, b' and `d$\pm$' form factors, and \
$m_{i}=\{m_{\sigma },\lambda _{\sigma }\}$ \ for the `c' contribution. The
limit $d \rightarrow 4$ is implied. The seemingly
divergent terms proportional to \ $\Gamma (2-d/2)$ are in fact finite due to
summations over $i$ (or $i, j$).

In terms of these formulas the one-loop EM form factors are written as
\begin{eqnarray}
 && F_{l}^p (q^{2})_{\pi, \sigma} = - F_{l}(q^2)_a +\frac{1}{2} F_{l}(q^2)_b +
  F_{l}(q^2)_c + F_{l}(q^2)_{d+} \,,
\nonumber \\
 && F_{l}^n (q^{2})_{\pi, \sigma} =  F_{l}(q^2)_a + F_{l}(q^2)_b + F_{l}(q^2)_{d-} \, ,
\;\;\;\;\;\;\;\;\;\;\;\;\;\; \quad \quad \quad (l=1,2) \,
\label{Fln}
\end{eqnarray}
for the proton and neutron.
Expressions (\ref{Eq:prot-neutr-kappa1}) - (\ref{cont+-}) of
subsect.~\ref{sec:one-loop} for AMMs follow from eqs.~(\ref{F2a}), (\ref{F2b}),
(\ref{F2c}) and (\ref{F2d+}) after putting $q^2$ equal to zero.

In order to calculate the electric and magnetic MSRs
in eqs.~(\ref{eq:radius-def}) we need, in addition to $F_2 (0)$, the
derivatives $F_1^{\prime} (0)$ and $G_M^\prime (0)$.
Calculating the derivatives of the form factors (\ref{F1a}) - (\ref{F2d+})
we obtain for the pion-loop diagram `a' in fig.~\ref{Fig.2}:
 \begin{eqnarray}
F_{1}^\prime (0)_{a} &=&-\frac{g^{2} \lambda_\pi^4}{8\pi ^{2} l_\pi^4 }
\int_{0}^{1}\int_{0}^{1}
\{ \frac{1}{D(m_\pi )} + \frac{1}{D(\lambda_\pi )} -
\frac{2}{ D(m_\pi ) +\beta (D(\lambda_\pi ) -D(m_\pi ) )}
 - 2 m_N^2 \alpha^2
\nonumber \\
&& \times \Big[ \frac{1}{D(m_\pi )^2 } + \frac{1}{D(\lambda_\pi )^2 } -
\frac{2}{ [D(m_\pi ) +\beta (D(\lambda_\pi ) -D(m_\pi ) ) ]^2 } \Big] \}
\nonumber \\
&& \times
(1-\alpha)^3 \beta (1-\beta)\  \mbox{\rm d}\alpha \mbox{\rm d}\beta \, ,
%%%\label{F1a-der}
\nonumber \\
 G_{M}^\prime (0)_{a} &=&-\frac{g^{2} \lambda_\pi^4}{8\pi ^{2} l_\pi^4 }
\int_{0}^{1}\int_{0}^{1}
\{ \frac{1}{D(m_\pi )} + \frac{1}{D(\lambda_\pi )} -
\frac{2}{ D(m_\pi ) +\beta (D(\lambda_\pi ) -D(m_\pi ) )}  \}
\nonumber \\
 && \times (1-\alpha)^3 \beta (1-\beta)\ \mbox{\rm d}\alpha  \mbox{\rm d}\beta \,,
\label{GMa-der}
\end{eqnarray}
for the diagram `b':
 \begin{eqnarray}
F_{1}^\prime (0)_{b} &=&\frac{g^{2} \lambda_\pi^4}{8\pi ^{2}  } \frac{1}{6}
\int_{0}^{1}
\{ 2 + m_N^2 \alpha^2  \Big[ \frac{1}{D(m_\pi )} + \frac{2}{D(\lambda_\pi )}
\Big] \}
\frac{\alpha^3 (1-\alpha )^2 }{D(m_\pi ) D(\lambda_\pi )^2 }\ \mbox{\rm d}\alpha \,,
%%%\label{F1b-der}
\nonumber \\
 G_{M}^\prime (0)_{b} &=&
\frac{g^{2} \lambda_\pi^4}{8\pi ^{2}  } \frac{1}{6} \int_{0}^{1}
\{ 2 - m_N^2 \alpha^2  \Big[ \frac{1}{D(m_\pi )} + \frac{2}{D(\lambda_\pi )}
\Big] \}
\frac{\alpha^3 (1-\alpha )^2 }{D(m_\pi ) D(\lambda_\pi )^2 }\  \mbox{\rm d}\alpha \,,
\label{GMb-der}
\end{eqnarray}
for the  $\sigma- $loop diagram `c':
  \begin{eqnarray}
F_{1}^\prime (0)_{c} &=&\frac{g^{2} \lambda_\sigma^4}{16\pi ^{2}  } \frac{1}{6}
\int_{0}^{1}
\{ 2 + m_N^2 (2-\alpha)^2  \Big[ \frac{1}{D(m_\sigma )} + \frac{2}{D(\lambda_\sigma )}
\Big] \}
\frac{\alpha^3 (1-\alpha )^2 }{D(m_\sigma ) D(\lambda_\sigma )^2 }
\ \mbox{\rm d}\alpha \,,
%%%\label{F1c-der}
\nonumber \\
 G_{M}^\prime (0)_{c} &=&
\frac{g^{2} \lambda_\sigma^4}{16\pi ^{2}  } \frac{1}{6} \int_{0}^{1}
\{ 2 + m_N^2 (4-\alpha^2) \Big[ \frac{1}{D(m_\sigma )} +\frac{2}{D(\lambda_\sigma )}
\Big] \} \frac{\alpha^3 (1-\alpha )^2 }{D(m_\sigma )
D(\lambda_\sigma )^2 }\ \mbox{\rm d}\alpha \,,
\label{GMc-der}
\end{eqnarray}
 and for the sea-gull diagram `d+':
 \begin{eqnarray}
F_{1}^\prime (0)_{d+} &=&
\frac{g^{2} \lambda_\pi^4}{4\pi ^{2} l_\pi^2 } \frac{1}{6}
\int_{0}^{1}\int_{0}^{1}
\{ \frac{1}{D(\lambda_\pi )^2} -
\frac{1}{[ D(m_\pi ) +\beta (D(\lambda_\pi ) -D(m_\pi ) )]^2} \nonumber \\
 &&- 4 m_N^2 \alpha^2  \Big[
 \frac{1}{D(\lambda_\pi )^3 } -
\frac{1}{ [D(m_\pi ) +\beta (D(\lambda_\pi ) -D(m_\pi ) ) ]^3 } \Big] \}
\nonumber \\
&& \times
(1-\alpha)^3 \beta^2 [3\alpha-2 +2(1-\alpha)\beta ] \  \mbox{\rm d}\alpha
\mbox{\rm d}\beta \, ,
%%%\label{F1d+-der}
\nonumber \\
 G_{M}^\prime (0)_{d+} &=&
 \frac{g^{2} \lambda_\pi^4}{4\pi ^{2} l_\pi^2 } \frac{1}{6}
\int_{0}^{1}\int_{0}^{1}
\{ \frac{1}{D(\lambda_\pi )^2} -
\frac{1}{[ D(m_\pi ) +\beta (D(\lambda_\pi ) -D(m_\pi ) )]^2}  \}
\nonumber \\
&& \times (1-\alpha)^3 \beta^2 [3\alpha-2 +2(1-\alpha)\beta ]\  \mbox{\rm d}\alpha
 \mbox{\rm d}\beta \,.
\label{GMd+-der}
\end{eqnarray}
The derivatives corresponding to the diagram `d-'
in fig.~\ref{Fig.2} are: $F_{1}^\prime (0)_{d-}= -F_{1}^\prime (0)_{d+}$  and
$G_{M}^\prime (0)_{d-}= -G_{M}^\prime (0)_{d+}.$
The functions $ D(m_{\pi,\sigma})$ and  $ D(\lambda_{\pi,\sigma})$
are defined in eqs.~(\ref{eq:Dml}).
The two-dimensional integrals in
eqs.~(\ref{GMa-der}) and (\ref{GMd+-der}) can be further reduced to the
one-dimensional ones after performing analytically the integration over $\beta$.

Finally, to check consistency of the model we need the wave-function
renormalization constant $Z_{2}$. If we write the nucleon self-energy in
eq.~(\ref{eq:self-energy}) as
\begin{equation}
\Sigma (p)= p\hspace{-0.5em}/ A(p^{2})+m_{N}B(p^{2}) \, ,
\end{equation}
then the constant reads
\begin{equation}
Z_{2}=\Big( 1-
\frac{\partial \Sigma(p)}{\partial p\hspace{-0.5em}/ } |_{p\hspace{-0.5em}/=m_N}
\Big)^{-1}=
\{1-A(m_{N}^{2})-2m_{N}^{2}[A^{\prime }(m_{N}^{2})+B^{\prime
}(m_{N}^{2})]\}^{-1}\,,
\label{eq:Z2}
\end{equation}
where $A^{\prime }(m_{N}^{2})$ denotes the derivative $dA(p^{2})/dp^{2}$
at $p^{2}=m_{N}^{2}$, and similarly for $B^{\prime }(m_{N}^{2}).$ Using
technique similar to that for the form factors one finds
\begin{eqnarray}
A(p^{2}) &=& \frac{g^{2}}{16\pi ^{2} }\int_{0}^{1}
\{ 3 \frac{\lambda_\pi^4}{l_\pi^4}  \Big[ \ln \frac{E(m_{\pi})}{E(\lambda _{\pi })}+
\frac{l_\pi^2 (1-\alpha ) }{E(\lambda _{\pi })}           \Big]
+ \frac{\lambda_\sigma^4}{l_\sigma^4} \Big[\ln \frac{%
E(m_{\sigma })}{E(\lambda _{\sigma })}
+ \frac{ l_\sigma^2 (1-\alpha) }{E(\lambda
_{\sigma })}\Big] \}      (1-\alpha)\   \mbox{\rm d}\alpha \,,
\nonumber \\
B(p^{2}) &=& \frac{g^{2}}{16\pi ^{2}}\int_{0}^{1}
\{ - 3 \frac{\lambda_\pi^4}{l_\pi^4} \Big[ \ln \frac{E(m_{\pi})}{E(\lambda _{\pi })}+
\frac{l_\pi^2 (1-\alpha ) }{E(\lambda _{\pi })}           \Big]
+ \frac{\lambda_\sigma^4}{l_\sigma^4}  \Big[\ln \frac{%
E(m_{\sigma })}{E(\lambda _{\sigma })}
+ \frac{ l_\sigma^2 (1-\alpha) }{E(\lambda
_{\sigma })}\Big] \}\ \mbox{\rm d}\alpha \,,
\label{eq:self-AB}
\end{eqnarray}
where $E(m_{\pi, \sigma})=D(m_{\pi, \sigma})+(m_N^2 -p^2) \alpha (1-\alpha)$
and \ $E(\lambda_{\pi, \sigma})
=D(\lambda_{\pi, \sigma})+(m_N^2 -p^2) \alpha (1-\alpha)$.
Due to the WT identity the constant $Z_{2}$,
calculated from eq.~(\ref{eq:Z2}), is equal to the vertex renormalization
constant $Z_{1}$ from subsect.~\ref{sec:renorm}. The latter in the present model is
\begin{eqnarray}
Z_1 = F_1^p (0)^{-1} = [ \ 1 - F_{1}(0)_{a}
+ \frac{1}{2} F_{1}(0)_{b} +F_{1}(0)_{c} + F_{1}(0)_{d+}\ ]^{-1} \,.
\label{Z1}
\end{eqnarray}

%%\end{appendix}

%%%%%%%%%%%%%%%%%%%%%%%%%%%%%%%%%%%%%%%%%%%%%%%%%

%%%%%%%%%%%%%%%%%%%    TABLES   %%%%%%%%%%%

\begin{table}[tbp]
\begin{center}
\begin{tabular}{|l|l|l|l|l||rrrr|}
Reference & $g_{\rho NN}\ (\kappa _{\rho })$ & $G_{\rho }$ & $g_{\omega
NN}\ (\kappa _{\omega })$ & $G_{\omega }$ & $R_{Ep}^{2}$ & $R_{Mp}^{2}$ & $%
R_{En}^{2}$ & $R_{Mn}^{2}$ \\ \hline
universality + SU(3)  & $\frac{1}{2} g_{\rho \gamma }\ (3.7)$ & $ \frac{1}{2%
}$ & $\frac{1}{2} g_{\omega \gamma }\ (0.)$ & $\frac{1}{2}$ & 0.39 & 0.40 &
$-$0.01 & 0.39 \\
effect. Lagrangian \cite{Noz90} & $2.66\ (3.7)$ & $0.52$ & $7.98\ (0.)$ & $0.47$ &
0.39 & 0.42 & $-$0.03 & 0.42 \\
effect. Lagrangian \cite{Feu99} & $2.07\ (3.01)$ & $0.41$ & $7.98\ (-0.12)$ & $%
0.47 $ & 0.34 & 0.29 & 0.02 & 0.26 \\
Bonn OBEP \cite{Mac87} & $2.66\ (3.7)$ & $0.52$ & $15.85\ (0.)$ & $0.93$ & 0.56
& 0.48 & 0.15 & 0.33
\end{tabular}
\end{center}
\caption{Vector-meson coupling parameters, ratios $G_V = g_{VNN}/g_{V\protect\gamma}$,
and the corresponding electric/magnetic mean-square radii
$R^2_{E/M} \equiv \langle r_{E/M}^2 \rangle_{\protect%
\omega, \protect\rho}$ (in fm$^2$) of the proton and neutron.}
\label{tab:VMD}
\end{table}

\begin{table}[]
\begin{center}
\begin{tabular}{|l|l|}
%% \hline
Subtractive & Multiplicative
\\ \hline
$F_{1}^{p}(q^{2})_{R}=F_{1}^{p}(q^{2})-F_{1}^{p}(0)+1 $ & $%
F_{1}^{p}(q^{2})_{R}=Z_{1}F_{1}^{p}(q^{2})$ \\
$F_{1}^{n}(q^{2})_{R}=F_{1}^{n}(q^{2})$ &
$ F_{1}^{n}(q^{2})_{R}=Z_{1}F_{1}^{n}(q^{2})$ \\
$F_{2}^{p/n}(q^{2})_{R}=F_{2}^{p/n}(q^{2})$ & $%
F_{2}^{p/n}(q^{2})_{R}=Z_{1}F_{2}^{p/n}(q^{2})$
\\
$\kappa_R = F_{2} (0) $ & $\kappa_R = Z_{1} F_{2}(0) $ \\
$ \langle r_{1}^{2} \rangle _{R} = 6 F_1^{\prime} (0) $ & $%
\langle r_{1}^{2} \rangle_{R} = 6 Z_{1} F_1^{\prime} (0) $ \\
$\langle r_{2}^{2} \rangle_{R} = 6 F_2^{\prime} (0) / F_2 (0) $ & $%
\langle r_{2}^{2}\rangle_{R} = 6 F_2^{\prime} (0) / F_2 (0)  $ \\
\end{tabular}
\end{center}
\caption{Renormalized EM form factors $F_{1,2} (q^{2})_{R}$,
anomalous magnetic moment $\kappa_R = F_{2} (0)_{R}$,
and mean-square radii $\langle r_{1}^{2} \rangle_R =6 F_1^{\prime} (0)_R, \
\langle r_{2}^{2} \rangle_R =6 F_2^{\prime} (0)_R / F_2 (0)_R$ in the
subtractive and multiplicative schemes
(formulas for AMM and MSRs are valid for the proton and neutron). }
\label{tab:renorm}
\end{table}

\begin{table}
\begin{center}
\begin{tabular}{|ccc|l|l|llll|}
%\hline
$\lambda _{\pi }$ (GeV) & $\lambda _{\sigma \text{ }}$(GeV) & $m_{\sigma}$ (MeV) & $%
\kappa _{p}$ & $\kappa _{n}$ & $R_{Ep}^{2}$ & $R_{Mp}^{2}$ & $R_{En}^{2}$ & $%
R_{Mn}^{2}$ \\
\hline
$\infty $ & $\infty $ & 915 & 1.57 & $-$3.52 & 0.82 & 0.68 & $-$0.34 & 0.57 \\
1.56 & $\infty $ & 915 & 1.793 & $-$1.913 & 0.78 & 0.65 & $-$0.28 & 0.70 \\
1.56 & 1.56 & 290 & 1.793 & $-$1.913 & 0.82 & 0.67 & $-$0.29 & 0.70 \\ \hline
& experiment &  & 1.792847337 & $-$1.91304272 & 0.74 & 0.74 & $-$0.119 & 0.77 \\
%\hline
\end{tabular}
\end{center}
\caption{Pion- and sigma-meson cut-off parameters, mass of the $\protect%
\sigma $, proton  and neutron anomalous magnetic moments (in nuclear
magnetons), total electric/magnetic mean-square radii
$R^2_{E/M} \equiv \langle r_{E/M}^2 \rangle $ (in fm$^2$).
Experimental values for AMMs are taken from \protect\cite{PDG99}, for
MSRs from \protect\cite{Eri88}.}
\label{tab:results}
\end{table}

\begin{table}
\begin{center}
\begin{tabular}{|c|rrrr|c|c|}
 $\langle r^2 \rangle$ (fm$^2$) & & $\pi,\sigma$ loops & & &
$\pi,\sigma$ loops + VMD   & experiment~\protect\cite{Eri88}   \\
 &  a  & a+b & a+b+c & a+b+c+d  & a+b+c+d+e & \\
\hline
 $ R_{Ep}^2$ &   0.24&   0.24 &   0.37 &   0.39  &   0.78
 &   0.74$\pm$0.02   \\
 $ R_{Mp}^2$ &   0.29 &   0.38 &   0.25 &   0.23  &   0.65
 &   0.74$\pm$0.11   \\
 $ R_{En}^2$ &$-$0.24 &$-$0.23 &$-$0.23 &$-$0.26  &$-$0.28
 &$-$0.119$\pm$0.004    \\
 $ R_{Mn}^2$ &   0.64 &   0.32 &   0.32 &   0.28  &   0.70
 &   0.77$\pm$0.13    \\
\end{tabular}
\end{center}
\caption{Contributions to the electric and magnetic mean-square radii from
the diagrams in fig.~\protect\ref{Fig.2}. The parameters are:
$\protect\lambda_{\protect\pi}=1.56$ GeV, $\protect\lambda_{\protect\sigma}
\rightarrow \infty$  and $m_{\protect\sigma}=915$ MeV for the $\protect\pi-,
\protect\sigma-$mesons,
and $ g_{\protect\omega NN}\ (\protect\kappa _{\protect\omega })=7.98\ (0.)$,
 $\ g_{\protect\rho NN}\ (\protect\kappa _{\protect\rho })=2.66\ (3.7)$
for the $\protect\omega-, \protect\rho-$mesons.}

\label{tab:radii}
\end{table}

%%%%%%%%%%%%%%%%%%%%  FIGURES  %%%%%%%%%%%%%%%%%%%%%%%%%%%%%%

\end{document}